\documentclass[twocolumn]{aastex62}
\usepackage{amsmath,amstext}
\usepackage[figure,figure*]{hypcap}
\usepackage{soul}
\usepackage{newtxmath} 
\usepackage{float}

\shorttitle{Milky Way Halo Age Gradient with BHB Photometry}
\shortauthors{Whitten et al.}

\begin{document}

\title{Constraints on the Galactic Inner Halo Assembly History 
from the Age Gradient of Blue Horizontal-Branch Stars}

\author[0000-0002-9594-6143]{Devin D.\ Whitten}
\affiliation{Department of Physics, University of Notre Dame, Notre Dame, IN 46556, USA}
\affiliation{JINA Center for the Evolution of the Elements, USA}

\author[0000-0003-4573-6233]{Timothy C.\ Beers}
\affiliation{Department of Physics, University of Notre Dame, Notre Dame, IN 46556, USA}
\affiliation{JINA Center for the Evolution of the Elements, USA}

\author[0000-0003-4479-1265]{Vinicius M.\ Placco}
\affiliation{Department of Physics, University of Notre Dame, Notre Dame, IN 46556, USA}
\affiliation{JINA Center for the Evolution of the Elements, USA}

\author[0000-0002-7529-1442]{Rafael M.\ Santucci}
\affiliation{Instituto de Estudos S\'ocio-Ambientais, Planet\'ario, Universidade Federal de Goi\'as, Goi\^ania, GO 74055-140, Brazil}
\affiliation{Instituto de F\'isica, Universidade Federal de Goi\'as, Campus Samambaia, Goi\^ania, GO 74001-970, Brazil}

\author{Pavel Denissenkov}
\affiliation{Department of Physics \& Astronomy, University of Victoria, Victoria, BC, V8W 2Y2, Canada}
\affiliation{JINA Center for the Evolution of the Elements, USA}

\author{Patricia B. Tissera}
\affiliation{Departamento de Ciencias Fisicas, Universidad Andres Bello, Av. Republica 220, Santiago, Chile}
\affiliation{Millennium Institute of Astrophysics, Av. Republica 220, Santiago, Chile}

\author{Andrea Mej\'{\i}as}
\affiliation{Departamento de Ciencias Fisicas, Universidad Andres Bello, Av. Republica 220, Santiago, Chile}
\affiliation{Millennium Institute of Astrophysics, Av. Republica 220, Santiago, Chile}

\author{Nina Hernitschek}
\affiliation{{Division of Physics, Mathematics and Astronomy, Caltech, Pasadena, CA 91125, USA}}

\author{Daniela Carollo}
\affiliation{INAF –-- Astrophysical Observatory of Turin, 10025 Torino, Italy}

\begin{abstract}
	
We present an analysis of the relative age distribution of the Milky Way halo, based on samples of blue horizontal-branch (BHB) stars obtained from the Panoramic Survey Telescope and Rapid Response System and \textit{Galaxy Evolution Explorer} photometry, as well a Sloan Digital Sky Survey spectroscopic sample. A machine-learning approach to the selection of BHB stars is developed, using support vector classification, with which we produce chronographic age maps of the Milky Way halo out to 40\,kpc from the Galactic center. We identify a characteristic break in the relative age profiles of our BHB samples, corresponding to a Galactocentric radius of $R_{\rm{GC}} \sim 14$\,kpc. Within the break radius, we find an age gradient of $-63.4 \pm 8.2$ Myr kpc$^{-1}$, which is significantly steeper than obtained by previous studies that did not discern between the inner- and outer-halo regions. The gradient in the relative age profile and the break radius signatures persist after correcting for the influence of metallicity on our spectroscopic calibration sample. We conclude that neither are due to the previously recognized metallicity gradient in the halo, as one passes from the inner-halo to the outer-halo region. Our results are consistent with a dissipational formation of the inner-halo population, involving a few relatively massive progenitor satellites, such as those proposed to account for the assembly of \textit{Gaia}-Enceladus, which then merged with the inner halo of the Milky Way.

\end{abstract}


\section{Introduction}

The $\Lambda$ cold dark matter cosmological paradigm describes the hierarchical growth of structure in the universe, where galaxies assemble via mergers of smaller, low-mass systems \citep{White:1978}. The continued process of hierarchical structure formation has largely been confirmed with the discovery of streams, tidal tails, over-densities, and numerous satellite galaxies of the Milky Way \citep{Majewski:2003, Belokurov:2006, Helmi:2008, Martin:2014, Shipp:2018}. However, a quantitative description of the assembly history of the Milky Way has eluded consensus.

The structural components of the Milky Way largely retain the signatures of their formation  \citep{Freeman:2002}, and thus present an opportunity to study the process of galaxy formation. The Milky Way halo is of particular importance, as the long dynamical times associated with this low-density component enable the persistence of substructures seen in various stages of their diffusion into the Galaxy. Low-mass stars in the halo can be nearly as old as the universe, while their spatial, kinematic, age, and chemical abundance distributions reflect their origins, whether that be in situ\footnote{In situ stars are taken to be those formed within the virial radius of the progenitor halo, either as a result of dynamical heating of an existing disc component or the transformation of gas brought in by satellite galaxies into stars.}, or in satellite galaxies accreted onto the primordial Milky Way.

Kinematical and chemodynamical studies of the halo have revealed it to comprise at least a dual system \citep[e.g.,][]{Gratton:2003, Carollo:2007, Miceli:2008, Nissen:2010, Carollo:2010, Beers:2012}, including a zero to mildly prograde net rotation inner halo with a peak metallicity of [Fe/H]$=-1.6$, and net retrograde outer halo with peak [Fe/H]$=-2.2$. Galaxy formation simulations predict stellar haloes to be formed mainly by the accretion of satellite galaxies, with contributions from in situ stars \citep{Zolotov:2009, Font:2011, Tissera:2012}. The properties of these accreted satellites would imprint features in the chemical abundances, age, and kinematics of the stellar populations in the stellar haloes \citep{Tissera:2013, Tissera:2014, Carollo:2018, Fattahi:2019, Fernandez:2019}. These features could be used to constrain the formation histories of the inner and outer regions of the stellar haloes. 


Following the second data release from \textit{Gaia} \citep{GAIA}, \citet{Belokurov:2018} demonstrated that halo stars of metallicity [Fe/H]$ > -1.7$ exhibit highly radial orbits (consistent with the claims of previous authors, e.g., \citealt{Chiba:2000}), suggesting a major accretion event by a massive ($10^{10}$\,$M_{\odot}$) satellite, between 8 and 11 Gyr ago. This progenitor, called the \textit{Gaia} Sausage \citep{Myeong:2018}, was confirmed by \citet{Helmi:2018}, whose findings suggested that the inner halo consists largely of debris from the accretion of a single progenitor, dubbed \textit{Gaia}-Enceladus, provided that \textit{Gaia}-Enceladus encompasses both the high eccentricity population from \citet{Belokurov:2018} and a retrograde component \citep{Koppelman:2018}. Using a sample of $\sim3000$ blue horizontal-branch (BHB) stars from the Sloan Digital Sky Survey (SDSS), \citet{Lancaster:2019} determined that this ancient structure constitutes at least $\sim50$\,\% of the metal-poor stellar halo within 30\,kpc, but acknowledged that it is unclear whether this structure is the residue of a single, two, or more radial infalls, as suggested by recent cosmological studies \citep{Kruijssen:2019}. \cite{Myeong:2019} provided the dynamical and chemical evidence of an additional accretion episode, distinct from Gaia-Enceladus. This satellite, referred to as the Sequoia galaxy \citep{Barba:2019}, contributed a stellar mass of $\sim5 \times 10^{7}$\,$M_{\odot}$ to the Milky Way, comparable to the Fornax dwarf spheroidal. The apparent complexity of the Milky Way assembly history leaves open the possibility that signatures of additional dwarf galaxy mergers may yet persist in the kinematics, chemical, or age distributions of the Milky Way's oldest stars.

While challenging to estimate, stellar ages are a powerful tool with which to constrain the merger history of the Galaxy. BHB stars have been successfully used to demonstrate a radial age gradient in the halo \citep{Preston:1991, Santucci:2015b, Carollo:2016, Das:2016}. Using UBV photometry for 4408 candidate field horizontal-branch stars, \citet{Preston:1991} demonstrated an increase in the $(B-V)_0$ color of $\sim0.025$\,mag with Galactocentric distance over $R_{G}=[2 ,12]$\,kpc, suggesting a decrease in the mean age of field horizontal-branch stars by a few Gyr. Following a rigorous identification of BHB stars from SDSS/SEGUE spectroscopy in \citet{Santucci:2015a}, \citet{Santucci:2015b} used de-reddened $(g-r)_0$ photometry of BHB stars to identify an increase in the mean colors of BHB stars from regions close to the Galactic center to $\sim 40$\,kpc, corresponding to an age difference of $2-2.5$\,Gyr, and produced an age map of the halo system up to $\sim25$\,kpc. In a later investigation, \citet{Carollo:2016} produced age maps up to $\sim 60$\,kpc by employing a large number of BHB stars selected on the basis of their colors from SDSS DR8 \citep{SDSS8}. Both works confirm the \citet{Preston:1991} result, and reveal the presence of numerous younger substructures in the outer-halo region. \citet{Carollo:2016} found a global age gradient of $-25.1 \pm 1.0$\,Myr kpc\,$^{-1}$, consistent with the result of $-30$\,Myr kpc$^{-1}$ by \citet{Das:2016}.


In a follow-up study of age gradients for Milky Way-mass galaxies simulated by the Aquarius Project, \citet{Carollo:2018} found an overall age gradient in the range of $[-8,-32]$\,Myr kpc\,$^{-1}$, for which the accreted component of the stellar population is largely responsible. These results suggest that the Milky Way formation history is dominated by the accretion of satellite galaxies with dynamical masses less than ${\sim}10^{9.5}M_{\odot}$. 

Using a suite of $N$-body simulations, \citet{Amorisco:2017} found that satellites that are accreted at higher redshift, and thus likely possess characteristically older stellar populations, deposit their material farther inside their host galaxies. Age gradients are thus a powerful diagnostic with which to probe the assembly history of the Milky Way. Unfortunately, the pioneering observational investigations were limited by sky coverage, sample size, and selection purity; more detailed studies of the nature of the observed gradient are required to distinguish between the widely varying age profiles seen in simulations of Milky Way-mass galaxies. 

Age gradients may also prove complementary to studies of the stellar density profile of the Milky Way, which was demonstrated by \citet{Pillepich:2014} to be a powerful diagnostic of a galaxy's accretion history, in addition to its total stellar mass. In contrast to the density profile of the dark matter halo, the stellar density profile is thought to follow an axisymmetric power law, with a distinct flattening in the inner-halo region, and a characteristic break radius occurring at $r \sim 25$\,kpc \citep[e.g.,][]{Saha:1985, Sesar:2011, Xue:2015, Hernitschek:2018}. The existence of a break radius in a halo's stellar density profile suggests the possibility of a similar break in its age profile.

In this work, we provide the first evidence for a characteristic break in the relative age profile of the Milky Way stellar halo, using a sample of BHB stars obtained from the Panoramic Rapid Response Survey Telescope (Pan-STARRS1) and the \textit{Galaxy Evolution Explorer} ($GALEX$). In Section 2, we discuss the various surveys used in the selection of BHB stars. In Section 3, we describe our selection strategy for BHB stars, based on a machine-learning photometric selection methodology. We describe the determination of relative ages for our BHB samples in Section 4, and model the chronographic distribution of these stars using maximum likelihood estimation (MLE). The results of our analysis of the radial age profile of the BHB samples are provided in Section 5. We discuss our interpretation of these results in Section 6, followed by concluding remarks in Section 7.

\begin{figure}
	\label{fig:usdss_nuv_cal}
	\begin{center}
		\includegraphics[trim = 0.20cm 0.50cm 0.5cm 0.00cm, clip, width=\columnwidth]{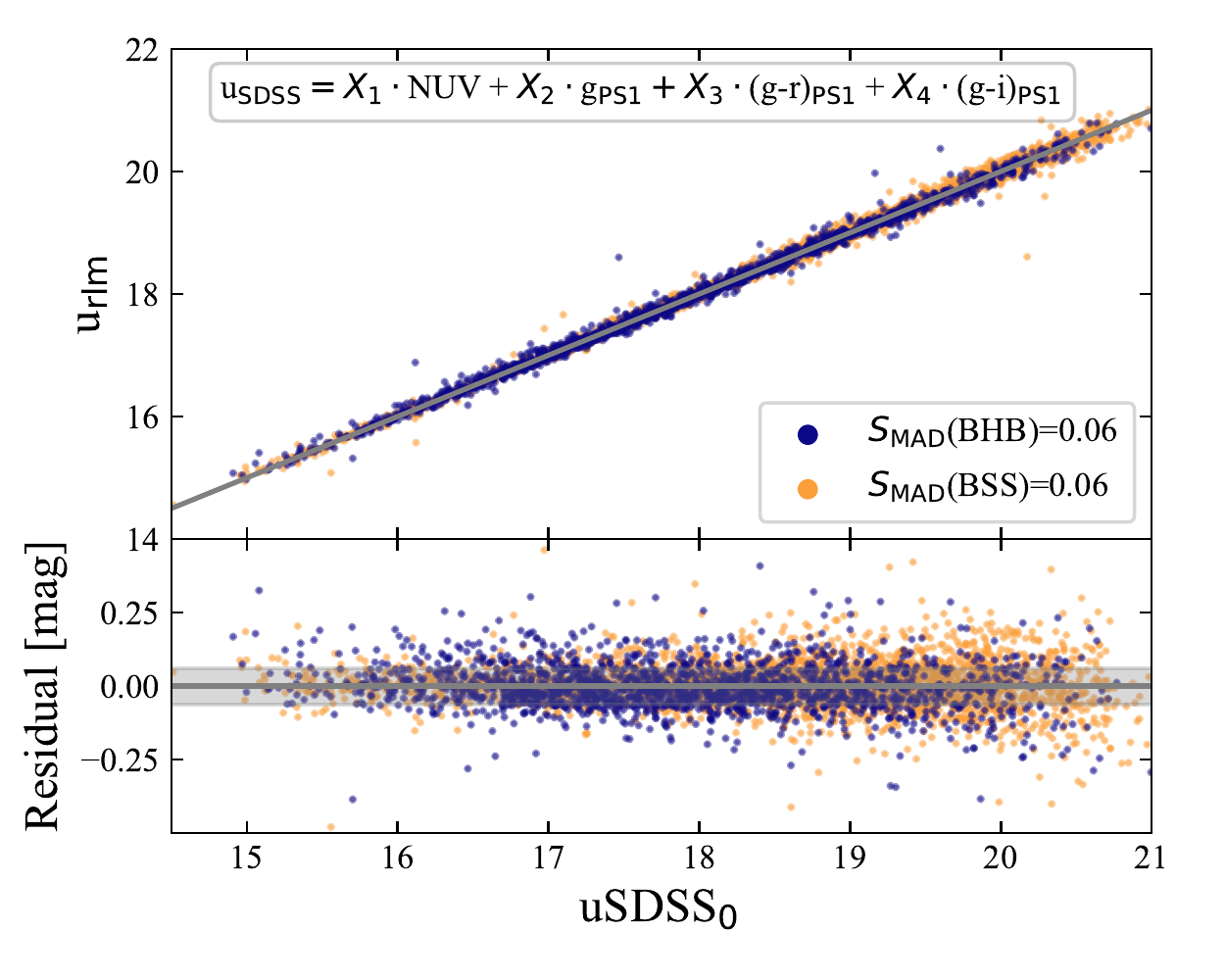}
		
		\caption{Upper panel: fit of the GALEX NUV and SDSS magnitudes and PS1 colors for calibration to $u_{\rm SDSS}$, according to the model uSDSS = $x_1\cdot$NUV + $x_2\cdot g_{\rm PS1} + x_3\cdot(g-r)_{\rm PS1}$ + $x_4\cdot(g-i)_{\rm PS1}$. Blue and orange dots indicate spectroscopically confirmed BHB and BSSs, respectively. Lower panel: residual plot of the resulting calibration.}
	\end{center}
\end{figure}

\section{Survey Samples}

In this section, we describe the photometric and spectroscopic survey catalogs used in the selection of halo BHB candidates in this work.

\subsection{SDSS}

We develop our selection methodology using the sample of spectroscopically verified BHB and blue straggler stars (BSSs) from SDSS/SEGUE described in \citet{Santucci:2015a}, hereby referred to as the SDSS spectroscopic sample. 
The sample consists of 4772 BHB and 7938 BSSs with medium-resolution ($R \sim 2000$) spectroscopy from SDSS DR8 \citep{Aihara:2011}, to a faint limit of $g\sim 21$. The spectroscopic criteria for the identification of BHB stars in this catalog utilized a number of properties in the stellar spectrum, including Balmer line widths and other gravity-sensitive features. We refer the interested reader to \citet{Santucci:2015a} for further details. A similar spectroscopic catalog was developed by \citet{Xue:2008}, however the \citet{Santucci:2015a} sample employed a somewhat larger ($u-g$) selection window, and contains a larger number of BHB stars. The $ugriz$ photometry was updated to the most recent version, SDSS DR12 \citep{Abolfathi:2018}.

\subsection{Pan-STARRS DR1}

The Pan-STARRS1 (PS1) survey is a set of high-cadence, multicolor, multi-epoch observations covering a large area of sky \citep{Tonry:2012}. The Pan-STARRS1 system is located on the island of Maui, Hawaii, and utilizes a 1.8\,m, $f$/4.4 telescope with a 1.4\,Gpix detector having a 3.3\,deg$^2$ field-of-view. The stacked PS1 3$\pi$ Steradian Survey \citep{Chambers:2016} observed the entire sky north of $\delta \sim -30^{\circ}$, in five bands ($g_{\rm P1}$, $r_{\rm P1}$, $i_{\rm P1}$, $z_{\rm P1}$, and $y_{\rm P1}$) to  limiting magnitudes of  23.3, 23.2, 23.1, 22.3, and 21.4, respectively. The \texttt{StackObjectThin} catalog was queried for objects with $-0.5 < (g-r)_{\rm{P1}} < 0.2$, and \texttt{primaryDetection}=1. A crude star-galaxy rejection, $\textrm{iPSFMag} - \textrm{iKronMag} \leq 0.06$\,mag, was employed to select for point-source objects \citep{Farrow:2014}. As recommended by \citet{Flewelling:2016}, \texttt{GaiaFrameCoordinates} were used for determination of positions, to ensure the highest quality astrometry possible.


\subsection{GALEX GUVCat}

The Galaxy Evolution Explorer \citep{Martin:2005} was the first far- and near-UV survey of the entire sky. The \textit{GALEX} instrument hosted a 50\,cm primary mirror, with a beam splitter for simultaneous broadband photometric measurements, in the far-UV ($\lambda_{\rm eff} \sim 1528$ \AA, 1344-1786 \AA) and near-UV ($\lambda_{\rm eff} \sim 2310$ \AA, hereafter NUV). We made use of the GALEX All-Sky Imaging survey, in particular the science-enhanced catalogs from GUVcat \citep{Bianci:2017}. This catalog provides a number of improvements to previous releases, including a 10\,\% larger sky coverage and removal of duplicate detections. We cross-matched the GALEX GUVCat with Pan-STARRS DR1, hereafter referred to as PS1xGALEX, with a search radius of 3\,arcsec, resulting in 1,098,309 unique sources.

\subsection{Color Transformations}

All catalogs were corrected for Galactic reddening and extinction according to \cite{Schlafly:2011}, where the $E(B-V)$ values included the 14\% recalibration of \cite{Schlegel:1998}, such that $E(B-V)=0.86 \cdot E(B-V)_{\textrm{1998}}$. Cuts in the point-spread function magnitude errors and the estimated extinction were then made, according to \texttt{PSFerr} $<$\,0.2\,mag, and $E(B-V) < 0.1$\,mag.

All photometric catalogs in this work were transformed to the corresponding SDSS photometry. For the PS1 catalog, this was done using the calibrations provided in \citet{Tonry:2012}. For the PS1xGALEX catalog, we developed the following transformation, $u_{\rm rlm}$, to approximate $u$SDSS based on the NUV and $g_{\rm PS1}$ magnitudes and $(g-r)_{\rm PS}$ and $(g-i)_{\rm PS1}$ colors, which resulted in the lowest scatter when compared to alternative color combinations

\begin{equation}
\begin{split}
u_{\rm SDSS} = x_1\cdot NUV + x_2\cdot g_{\rm PS1} + x_3\cdot (g-r)_{\rm PS1} + \\ x_4\cdot(g-i)_{\rm PS1}.
\end{split}
\end{equation}

The resulting calibration is shown in Figure~\ref{fig:usdss_nuv_cal} for the subset of the SDSS spectroscopic sample with SDSS, Pan-STARRS, and GALEX NUV photometry. The optimal set of coefficients were found to be:  $X_1=0.22, X_2 = 0.78, X_3 = -0.21, X_4 = -0.14$. Scatter in the calibration is estimated with the median absolute deviation, scaled to the corresponding standard deviation, $S_{\rm MAD}$, for which we, obtain $S_{\rm MAD}(u_{SDSS}) < 0.06$\,mag.




\section{BHB Candidate Selection}\label{Section:BHB_Selection}

\begin{figure*}
	\label{fig:SVC_optimization}
	\begin{center}
		\includegraphics[trim = 1cm 0.00cm 0.75cm 0.50cm, clip, width=\textwidth]{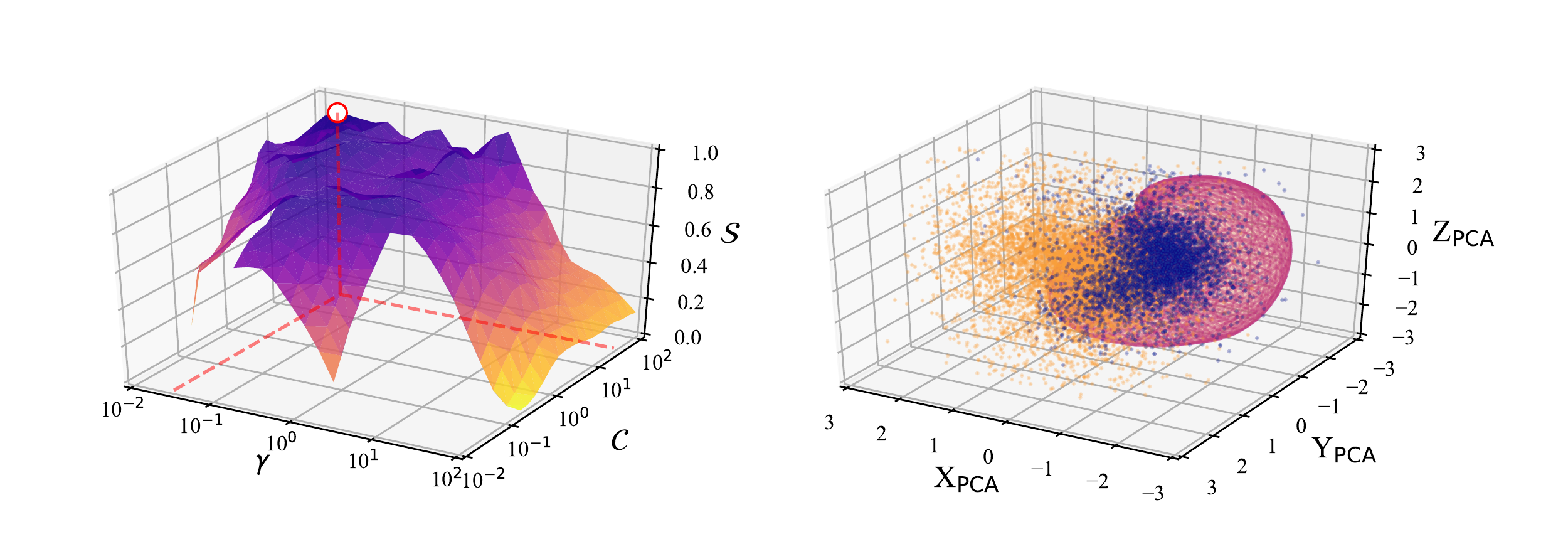}
		
		\caption{Left panel: hyperparameter  optimization for the support vector classifier. Optimal parameters $(C, \gamma) = (31.62, 0.316)$ were chosen where the score, $\mathcal{S}$, is maximized.
		Right panel: BHB selection function corresponding to the optimized SVC, in the 3D photometric PCA space. Blue and orange dots denote spectroscopically confirmed BHB stars and BSSs, respectively.}
	\end{center}
\end{figure*}

Two unique photometric selections of BHB stars are performed in this work. Although we employ the $(i-z)$ selection developed in \citet{Vickers:2012} as a verification of our selection methodology, we demonstrate below that we can obtain a higher sample purity with our new selection method, which we develop with the SDSS spectroscopic sample of \citet{Santucci:2015b}, described previously.

\subsection{Support Vector Classification}

Two gravity-sensitive colors were used for selection of BHB stars, in order to distinguish between the otherwise higher surface-gravity foreground A-type contaminants -- referred to as BSSs in this work -- in the initial $(g-r)_0$ selection window. The region of the Balmer break has most often been exploited \citep{Yanny:2000,Sirko:2004, Carollo:2016, Thomas:2018}, where a broadening of the Balmer lines is seen for high-gravity stars due to Stark pressure broadening. Additionally, \citet{Vickers:2012, Vickers:2014} demonstrated the effectiveness of the Paschen break ($8200 $\AA) as a surface gravity-dependent feature detectable with photometry, although perhaps to a lesser degree than the Balmer break.

We first perform principal component analysis (PCA; \citealt{PCA}) with the inputs $(u-g)_0$, $(g-r)_0$, and $(i-z)_0$ of the SDSS spectroscopic sample.
 The first and second principal component captured 45\% and 41\% of the variance, respectively, while the third component captured 14\%.  We therefore retain all three principal components throughout the selection procedure, described in the following section. An equivalent transformation is performed on all photometric catalogs, using the principal components of the SDSS spectroscopic sample.

We develop the 3D BHB selection function using support vector classification (SVC; \citealt{Boser:1992}, \citealt{Pedregosa:2011}). This supervised learning model is a binary classifier which, given data of distinct classes, seeks to find a transformation such that the data classes become maximally separable by a hyperplane in a corresponding feature space. This feature space is dictated by the selected kernel function, the weights of which are optimized during the training procedure. Training of an SVC is therefore concerned with the determination of the appropriate transformation to be applied to the data inputs, such that this separation is effective.
To accomplish this, our SVC employs a Gaussian radial basis function (RBF) as a kernel, with which the photometric inputs are transformed. 
Like many machine-learning algorithms, an SVC with an RBF kernel makes use of two hyperparameters, which must be tuned, in addition to the supervised training process. The first hyperparameter, known as the regularization parameter, $C$, governs the extent to which misclassifications are penalized during training. The second hyperparameter is the width of the Gaussian RBF, $\gamma$, which controls the influence of data far from the classification boundary. 
We optimize over the hyperparameters, $C$ and $\gamma$, using the SDSS spectroscopic sample, which we split into training and validation sets of 65\% and 35\%, respectively. The validation fraction of 35\% is somewhat higher than typically advised. This fraction was chosen to ensure that the validation sample was sufficiently large to study the affect of the apparent magnitude on the classification purity and recovery fractions. 

We compose a grid of SVCs across the hyperparameter range $10^{-2} < C < 10^{2}$ and $10^{-2} < \gamma < 10^{2}$, and track the performance of each SVC in terms of the resulting BHB classification purity, $f_C$, and recovery fraction, $f_R$. SVC over-training can result in an over-specified decision function and isolated `decision pockets' around individual data in the training set.  To check for this, we additionally track the inverse variances of $f_C$ and $f_R$, which are determined using 30 randomly selected subsamples of the validation set from the SDSS spectroscopic sample. Each subset results in differing classification and recovery fractions, as stars fall into the decision pockets, evidenced by an increase in the classification and recovery variances. 
We therefore introduce a score parameter for the hyperparameter optimization, $\mathcal{S}$, which incorporates the classification and recovery fractions, as well as the scale estimates of these values, as determined from the 30 iterative resamples of the data:

\begin{equation}\label{score_eq}
\mathcal{S}(\gamma, \mathcal{C}) =\frac{ f_C(\gamma, \mathcal{C}) \cdot f_R(\gamma, \mathcal{C})}{S_{\rm MAD}(f_C(\gamma, \mathcal{C})) \cdot S_{\rm MAD}(f_R(\gamma, \mathcal{C}))}
\end{equation}

Here, $S_{\rm MAD}$ denotes the estimate of scale, the median absolute deviation. The result of the grid optimization is shown in the left panel Figure~\ref{fig:SVC_optimization}, where we take the combination of $(C, \gamma)$ which maximize Eq.~\ref{score_eq}. These were found to be $(C, \gamma) = (31.62, 0.316)$, for which the resulting SVC decision boundary is shown in the right panel of Figure~\ref{fig:SVC_optimization}.

		

\begin{figure*}
	\label{fig:SVC_purity}
	\begin{center}
		\includegraphics[trim = 0.50cm 0.50cm 0.75cm 0.50cm, clip, width=\textwidth]{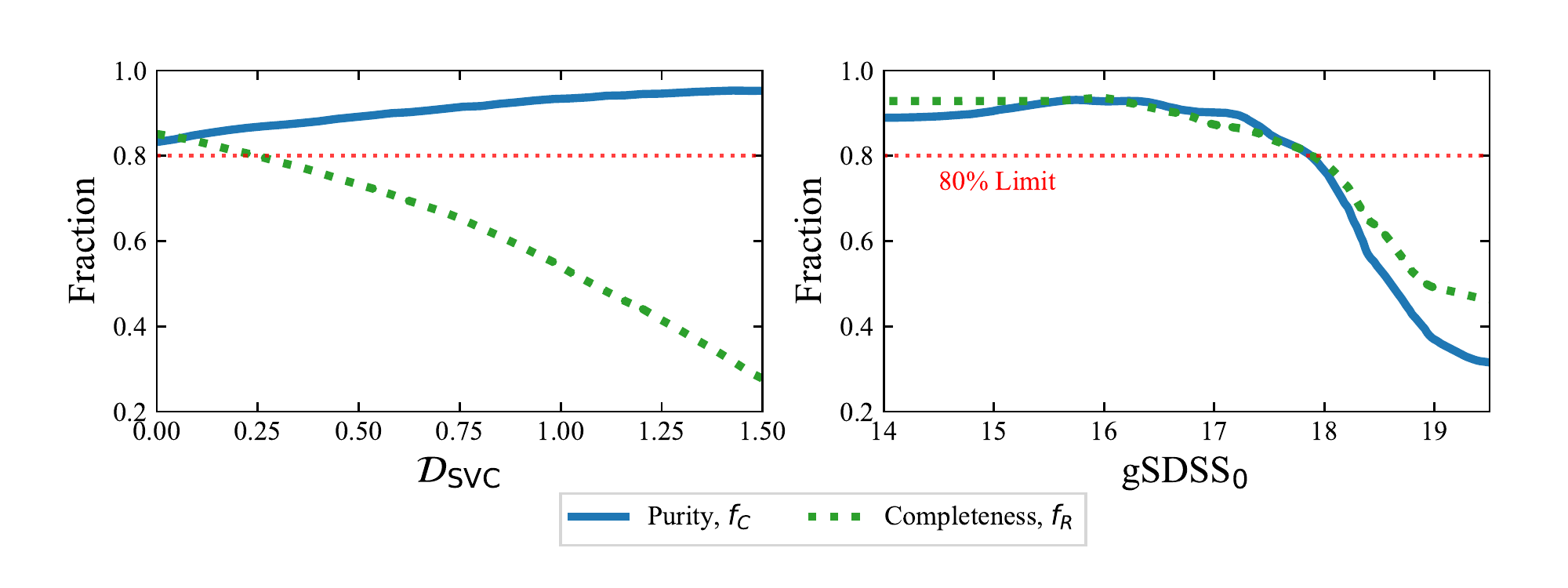}
		
		\caption{Left panel: BHB purity and completeness of the SVC selection as a function of hyperplane distance, $\mathcal{D}$. Right panel: purity and sample completeness as a function of apparent magnitude.}
	\end{center}
\end{figure*}

The resulting decision function of the SVC provides the distance, $\mathcal{D}$, of each star from the hyperplane. While this value is not strictly a probability, the sign and magnitude of $\mathcal{D}$ can be interpreted as a measure of confidence in the resulting classification. We explore the influence of the hyperplane distance on the resulting purity and completeness of the SVC in the left panel of Figure~\ref{fig:SVC_purity}. A marginal increase in the purity is seen with a higher restriction on the hyperplane distance, $\mathcal{D}$, while the completeness of the selection decreases precipitously. We therefore define our selection function for BHB stars in the PS1xGALEX catalog as the SVC decision boundary at $\mathcal{D}=0$.  
Additionally, we investigate in Figure~\ref{fig:SVC_purity} the purity and completeness of our selection function with apparent magnitude. To maintain a purity of $f_C > 80$\,\% we implement a magnitude limit of $g_{\rm SDSS, 0} < 18.1$ throughout the rest of our work. Finally, we subject the PS1xGALEX catalog to the SVC selection procedure, which resulted in $14,647$ BHB candidates. Hereafter, we refer to this selection as the PS1xGALEX-SVC sample.

\subsection{The \citet{Vickers:2012} Selection}

As a validation of the SVC selection methodology, we additionally perform the BHB selection procedure outlined by \citet{Vickers:2012} with the PS1 catalog. This method relies on the surface-gravity sensitivity of the Paschen region of the spectral energy distribution, corresponding to the $z$-band. This selection is defined as follows:

\begin{equation}
-0.3 < (g-r)_0 < 0.0
\end{equation}

\begin{equation}
 0.5\left[(g-r)_0 + 0.3\right] - 0.20 < (i-z)_0 < -0.03
\end{equation}


The purity of classification was determined by \citet{Vickers:2012} to be ${\sim}74$\,\%, in part because of the limited sensitivity with which this feature can be used to distinguish between the more numerous foreground A-type stars. This selection hereby referred to as the PS1-$(i-z)$ sample, resulted in $20,351$ stars.

\subsection{Quasar Contamination}

Within the faint limit of our samples, $g<18.1$, we expect a quasar density of $n(g < 18.1) \sim 0.1$\,deg$^{-2}$\,mag$^{-1}$ \citep{Pei:1995}, corresponding to a total of $N\sim10^3$, when taking into account the footprint of Pan-STARRS DR1. Considering the stringent color selections of BHBs from our catalogs, this number is effectively an upper limit, as we expect far less contamination in our SVC selections. 
We test this expectation by matching the PS1 and PS1xGALEX BHB catalogs with the Large Quasar Astrometric Catalogue \citep{Gattano:2018}, a catalog consisting of nearly all known quasars ($443,725$ sources), including new identifications from the DR12Q release of SDSS and \textit{GAIA} DR1. With a 5\,arcsec search radius, we find 22 unique matches for the PS1xGALEX-SVC sample, and 3380 for the PS1-$(i-z)$ sample. These sources were excluded from our analysis.

\subsection{Disk Contamination}

Using stellar density models for the thick disk and halo from \citet{Mateu:2018}, we estimate the ratio of thick disk to halo stars at $|Z_{G}| = 2.5$\,kpc to be $\sim11$\%. This is certainly an overestimation for BHB stars. For a disk star to pass our preliminary $g-r$ selection of $-0.5 < g-r < 0.2$, it would necessarily be hot, $T_{\rm eff}$ = 7500 - 9500 K. Stars in this effective temperature range consist of high-surface gravity dwarfs and horizontal-branch stars. These high-surface gravity A-type stars are precisely the type that the photometric selection was designed to remove, on the basis of surface gravity indicators. They differ by at least an order of magnitude in surface gravity from horizontal-branch stars. While it is feasible that horizontal-branch stars do exist in the thick disk, as seen in \citet{Bensby:2013}, the bulk of giant stars in the thick disk possess significantly higher metallicities than what is thought to be able to  produce a BHB star. For the metallicity ranges of the thin and thick disk, these would be red giant clump stars instead, and therefore would not pass our $g-r$ color selection.

\section{Methodology}
In this section, we discuss the determination of astronometry and photometric age estimates for the BHB samples. 

\subsection{Astrometry}
Distances  for stars in the BHB samples are derived from the line-of-sight distance modulus, using the absolute magnitude calibration developed by \citet{Deason:2011}. We calculate $M_g$ as 

\begin{equation}
\begin{split}
M_g = 0.434 - 0.169\cdot(g-r)_0 + 2.319\cdot(g-r)_{0}^2 + \\ 20.449\cdot(g-r)_{0}^3 + 94.517\cdot(g-r)_{0}^4 .
\end{split}
\end{equation}

Scatter in the absolute magnitude calibration was determined to be less than 0.1\,mag, corresponding to a distance uncertainty of ${\sim}5-10$\,\%. However, the absolute magnitude for BSSs is ${\sim}2$\,mag fainter, thus the uncertainty in our distance estimate is primarily determined by our sample purity. For our SVC sample purity of 80\,\%, the uncertainty in the distance estimates is then ${\sim}20$\,\%. 

Together with the equatorial coordinates, ($\alpha$, $\delta$), we compute the Galactocentric Cartesian coordinates, $X_{\rm G}$, $Y_{\rm G}$, $Z_{\rm G}$, and Galactic latitude, $l_{\rm G}$ and $b_{\rm G}$, assuming a distance of 8.5\,kpc of the Sun from the Galactic center. We then compute the Galactocentric distance, $R_{G}$, defined as $R_{G} = \sqrt{X_{\rm G}^2 + Y_{\rm G}^2 + Z_{\rm G}^2}$. For all samples, we employ a cut of $|Z_{G}| > 2.5$\,kpc and $|b_{G}| > 15^{\circ}$, to reduce contamination from disk-system stars, with an additional minimum Galactocentric radius of $R_{G} > 5$\,kpc.

\subsection{Chronography}

Age estimates for stars in the BHB samples considered in this work are made using the horizontal-branch population synthesis tool of \citet{Denissenkov:2017}, to which we refer the interested reader for details. Models were generated using revision 7624 of the MESA stellar evolution code \citep{MESA:2011, MESA:2013}, in which solar chemical abundances from \citet{Asplund:2009} were adopted. A model grid was constructed, taking into account age, mean $(g-r)_0$ color, and metallicity, which can then be interpolated to estimate an age given $(g-r)_0$ and [Fe/H]. For the SDSS spectroscopic sample, we use the estimates of [Fe/H] produced by the SEGUE Stellar Parameter Pipeline (SSPP; \citealt{Lee:2008a, Lee:2008b,SEGUEc}). For all other catalogs, we adopt the SDSS sample median [Fe/H] $= -1.75$.

\subsection{The Dependence of BHB Age Estimates on [Fe/H]}

We first investigate the influence of individual [Fe/H] estimates on the inferred age of our BHB stars. To do so, two distinct estimates of age are made for the SDSS spectroscopic sample, first using the individual [Fe/H] estimates from the SSPP (hereby referred to as the corrected age estimate, $A_{\mathrm{[Fe/H]}}$), and secondly by adopting the median [Fe/H] $= -1.75$ for all stars in our sample (hereafter referred to as the photometric age estimate, $A_{(g-r)_0}$).
\begin{figure}
	\includegraphics[trim = 0.0cm 0.50cm 0.75cm 1.5cm,clip, width=\columnwidth]{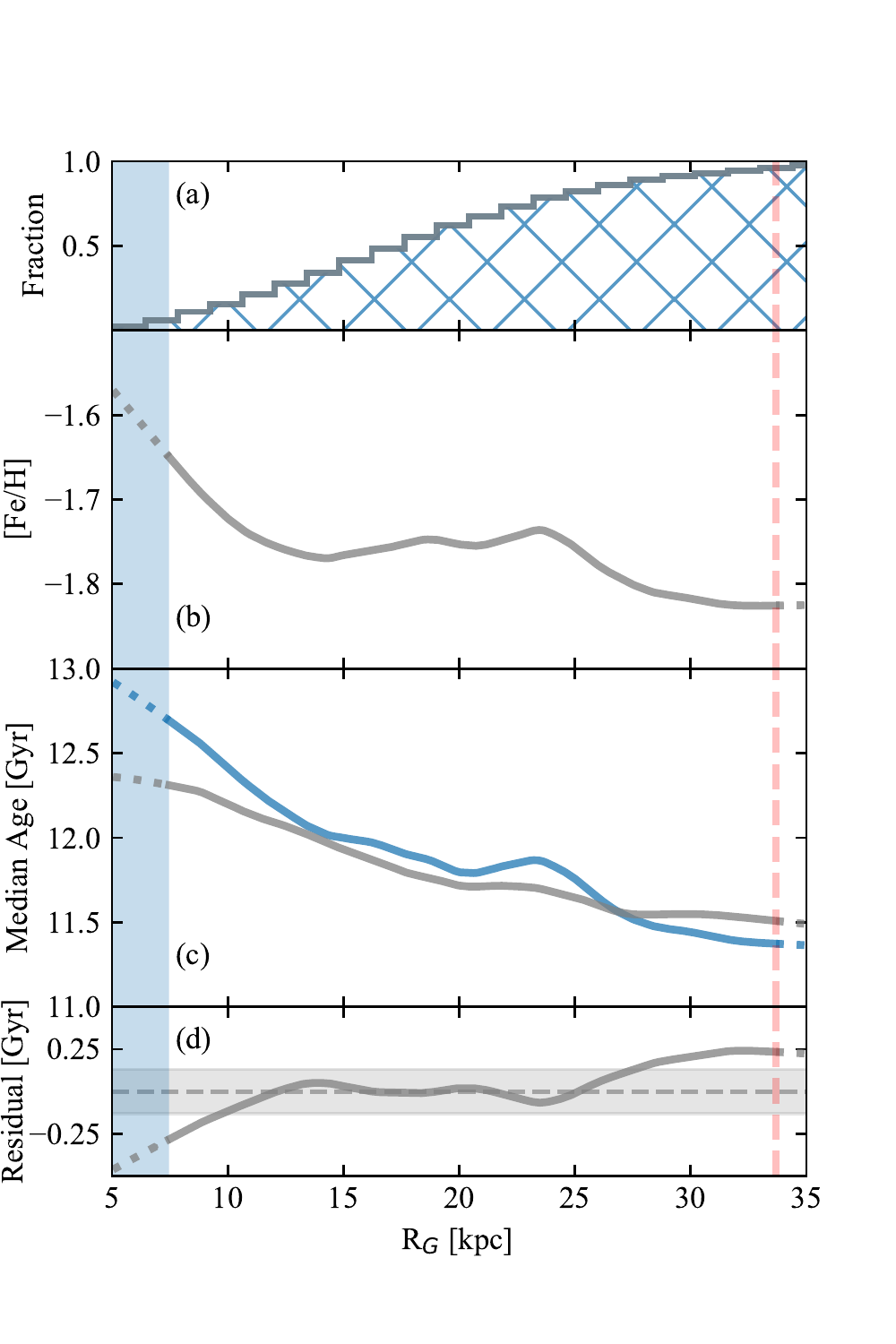}
	\caption{Panel (a): cumulative distribution of BHB stars from the SDSS spectroscopic sample with respect to Galactocentric distance. 
    Panel (b): median metallicity of the SDSS spectroscopic sample with respect to Galactocentric distance.
	Panel (c):	age distributions of the SDSS spectroscopic sample from \citet{Santucci:2015a}, for both photometric (gray) and corrected age estimates from \citet{Denissenkov:2017} (blue). 
	Panel (d): difference between the photometric and corrected age profiles. The broken red line represents the radius beyond which the number of stars decreases below 5\,\%.}
	\label{fig:santucci_offset}
\end{figure}

We use locally weighted scatterplot smoothing (LOWESS; \citealt{LOWESS}) for all profiles visualized in this work, for which we set the local sampling fraction to 25\,\% throughout.
Panel (a) of Figure~\ref{fig:santucci_offset} shows the cumulative radial distribution of the SDSS spectroscopic sample, and panel (b) shows the median metallicity profile of the sample. In panel (c) of Figure~\ref{fig:santucci_offset}, we compare the age estimate obtained from the $(g-r)_0$ photometry to the age estimate corrected for metallicity. A slight offset, of order $A_{(g-r)_0} - A_{\mathrm{[Fe/H]}} = -100$\,Myr, is seen between the profiles. We subtract this offset and plot the resulting residual profile in panel (d). As we are not primarily interested in the absolute ages, we correct for this offset in the photometric age estimate, and find a standard deviation in the residuals of $\sigma=130$\,Myr. Within the Galactocentric radius range of $10 < R_{G} < 27$\,kpc, the LOWESS regression of the corrected residual is within $1\sigma$. We can then assume that, within this range, the photometric age estimate is representative, and we apply the photometric age estimates assuming a median [Fe/H] $ = -1.75$ for all of the photometric-only BHB catalogs in this work. For further comparison, projected $X_{G}$ vs. $Z_{G}$ age distributions for both age estimate techniques are provided in Figure~\ref{fig:santucci_map} of the Appendix. 

\subsection{Radial Age Profiles}
\begin{figure}
	\includegraphics[trim = 0.0cm 0.50cm 0.0cm 0.0cm, width=\columnwidth]{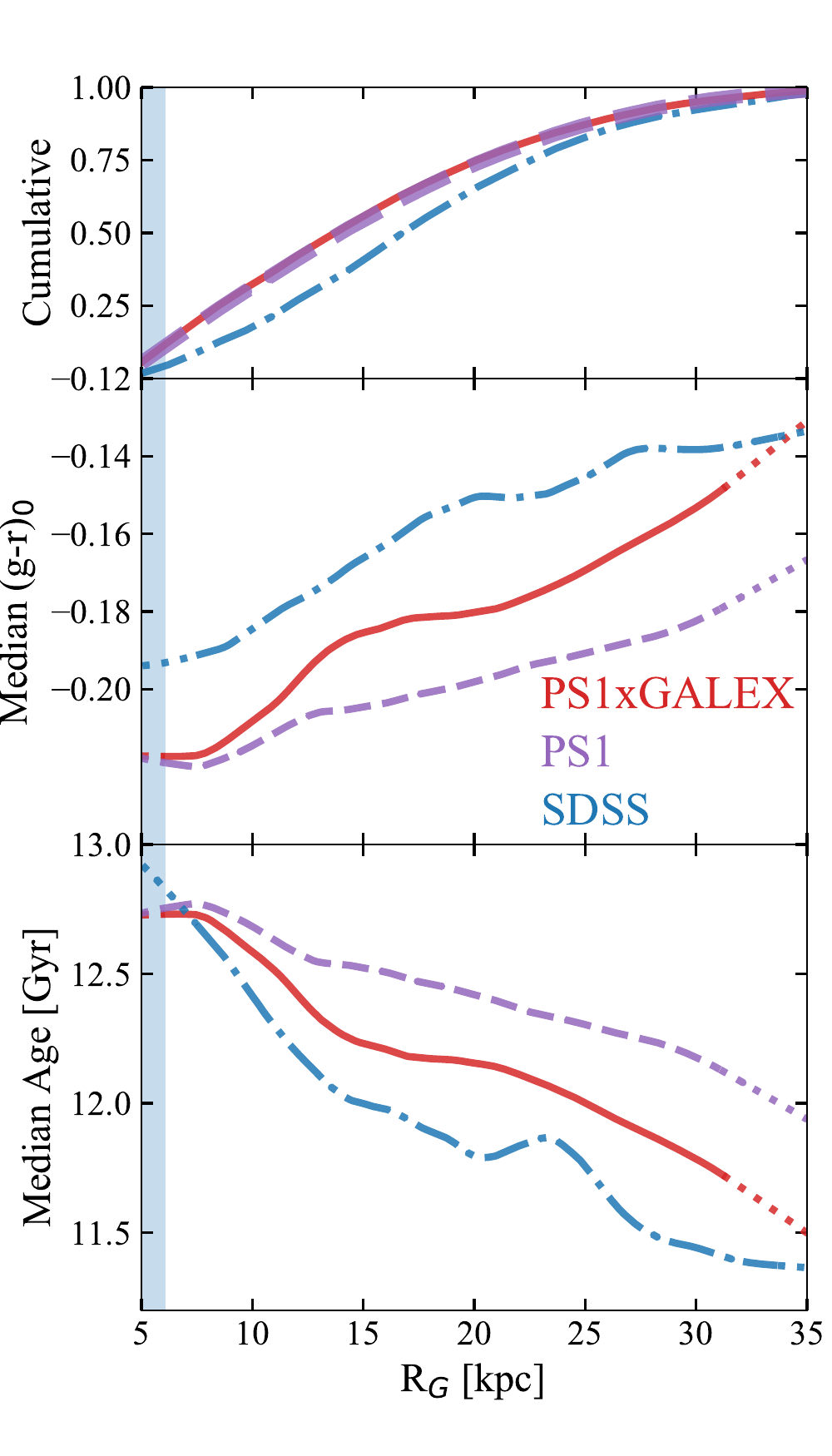}
	\caption{Top Panel: cumulative distributions of the PS1xGALEX-SVC (red, solid), PS1-$(i-z)$ (violet, dashed), and SDSS spectroscopic (blue, dashed-dotted) samples, with respect to Galactocentric distance.
	Center panel: median $(g-r)_0$ color of the BHB samples as a function of Galactocentric distance.
	Bottom Panel: median age of the BHB samples as a function of Galactocentric distance.}
	\label{fig:radial_profile}
\end{figure}

The cumulative radial distribution for each BHB sample is shown in the top panel of Figure~\ref{fig:radial_profile}, followed by the median $(g-r)_0$ and age profiles in the center and bottom panels, respectively. We again use LOWESS regression to visualize the profiles, with the local sampling fraction set to 25\%. We mark the radial thresholds for each sample beyond which the star count drops below 5\% of the sample, 29.7, 31, and 33.6\,kpc for the PS1xGALEX-SVC, PS1-$(i-z)$, and SDSS spectroscopic catalogs, respectively.


\begin{figure*}
	\includegraphics[scale=0.85, trim =10.0cm 8.50cm 11.00cm 4.0cm, clip, width=\textwidth]{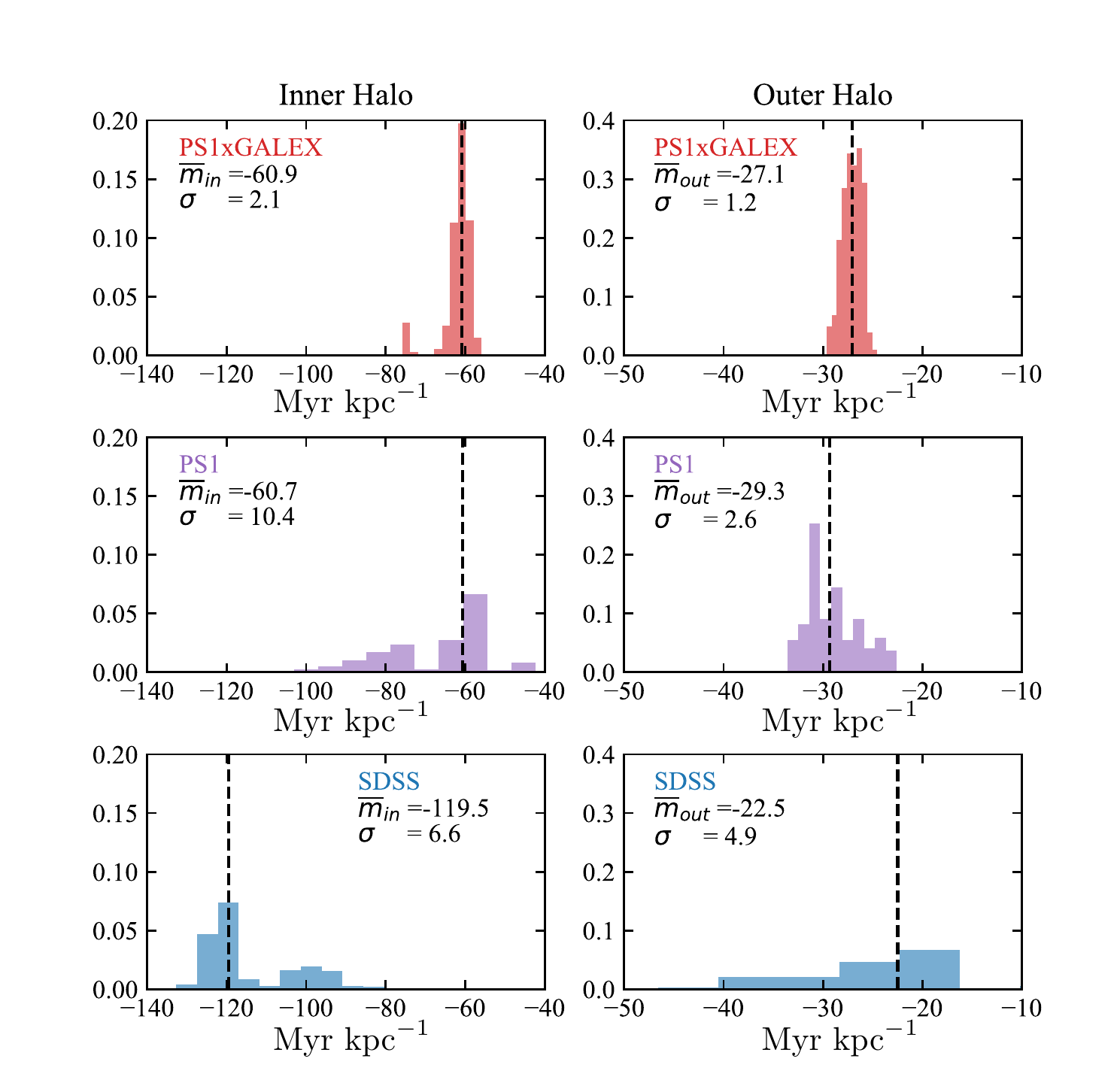}
	\caption{Distributions of Galactocentric radial age gradient estimates of the inner-halo (left panels) and outer-halo (right panels) regions. Gradients were determined for the PS1xGALEX (top), PS1-$(i-z)$ (middle), and SDSS spectroscopic (bottom) samples from 500 iterations of RANSAC regression, from which the mean $\overline{m}$ and standard deviation $\sigma$ estimates are shown and listed in Table.~\ref{table:param_values}.}
    \label{fig:ransac_plot}
\end{figure*}


We investigate the presence of a break radius in our BHB samples using maximum likelihood estimation. We begin with a segmented linear model, defined as follows:

\begin{equation}\label{seg_regression_eq}
\begin{split}
A(R_{G}) = A_0 - m_{\rm i}\cdot R_{G} + \\ (m_{\rm out} - m_{\rm in})\cdot(R_{G}-R_{b})\cdot H(R_{G} - R_{b})
\end{split}
\end{equation}

\noindent Here, $A_{0}$ is the intercept -- the age at $R_{G}=0$ -- and $m_{in}$ and $m_{out}$ are the age gradients of the inner and outer linear profile, respectively, delineated by the break radius, $R_b$. We hereafter refer to the inner- and outer-halo regions, distinguished on the basis of the break radius, as the IHR and OHR, respectively. The Heaviside function, $H$, acts to switch from the IHR to OHR slope at the break radius. To build our likelihood function, we assume that the data are distributed about the segmented linear model fit according to a Gaussian, for which we assign separate standard deviations, $\sigma_{\rm in}$ and $\sigma_{\rm out}$, for the inner and outer profiles, respectively. The resulting probability function is

\begin{equation}
\rho(A_i | R_{G,i} ; \theta) = \frac{1}{\sqrt{2 \pi \sigma^2(R_{G, i})}} \mathrm{exp}\left[\frac{-(A_i - A(R_{G,i} | \theta))^2}{ 2 \pi \sigma^2(R_{G,i})}\right]
\end{equation}

\noindent Here, $\sigma(R_G)$ is simply a step function:

\begin{equation}
\sigma(R_G) = 
\begin{cases}
\sigma_{\rm in}, & \text{if } R_G \le R_b,\\
\sigma_{\rm out}, & \text{if } R_G> R_b.
\end{cases}
\end{equation}

The model parameters, which we designate with $\theta$, are then $A_0$, $m_{\rm in}$, $m_{\rm out}$, $R_b$, $\sigma_{\rm in}$, and $\sigma_{\rm out}$. Our set of input parameters, $\mathcal{X}$, are simply the age, $A_i$, and Galactocentric radius, $R_{G,i}$. The optimal parameters are then determined using MLE. The corresponding log-likelihood function is:

\begin{equation}
\begin{split}
\ln{\mathcal{L}(\theta | \mathcal{X})} = \sum_{i}\ln{\rho(\mathcal{X}_i | \theta)} \\ 
= -\sum_{i}^{N}\ln{\sigma(R_{G,i})} - \frac{1}{2}\sum_{i}^{N} \left( \frac{ A_i - A(R_{G, i} | \theta)}{\sigma(R_{G,i})} \right)^{2}.
\end{split}
\end{equation}

It is important that the segmented linear model in which a break radius is assumed be compared to a simple, no-break, linear model. In this case, the log-likelihood model is essentially identical, where we have substituted the segmented linear model of Eq.~\ref{seg_regression_eq} with a simple linear model. We discuss the goodness-of-fit comparisons in Section~\ref{results_section}.

We determine best-fit parameters for our likelihood functions by sampling over the parameter space using the Python module \texttt{emcee} \citep{emcee} implementation of Goodman \& Weare's affine invariant Markov Chain Monte Carlo routine (MCMC; \citealt{Goodman:2010}). For all samples, all model priors are taken to be uniform. We additionally assert that the intercept of our segmented and simple linear models, $A_0$, not exceed $13.8$\,Gyr, and that the standard deviations, $\sigma_{\rm in}$ and $\sigma_{\rm out}$ be positive. For the segmented linear model, we set the edges of the uniform prior to be at the minimum and maximum Galactocentric radii present in the sample. For the SDSS spectroscopic sample, however, we limit the break radius to $R_b < 22$\,kpc, to avoid the otherwise uninteresting deviation seen to occur at $\sim23$\,kpc in the LOWESS age profile in Figure~\ref{fig:radial_profile}. While real, this deviation is likely an artifact of the small footprint of the SDSS spectroscopic sample, and thus is not representative of the underlying age profile.

\subsection{Random Sample Concensus}
Additionally, we estimate the slopes of the IHR and OHR profiles using a random sample concensus approach (RANSAC; \citealt{RANSAC}). This nondeterministic algorithm achieves a consensus on the optimal linear model by iterative sampling of the dataset, effectively mitigating outliers. First, we split the BHB samples by Galactocentric radius at an initial value of $R_b = 15.0$\,kpc, and evaluate the gradient in each region separately. We then iterate the RANSAC linear parameter determination over 500 resamples of each region. In each iteration, we infer a break radius from the intersection of inner and outer age profiles. From these 500 estimates of $m_{in}$, $m_{out}$, and $R_b$, we determine the median value and standard deviation, the distributions of which are shown in Figure~\ref{fig:ransac_plot}.



\clearpage
\section{Results and Discussion}\label{results_section}

The parameters of the segmented linear model are determined for the Galactocentric radial age profiles of the three BHB catalogs described above, using MLE implemented with an MCMC routine. The resulting posterior distributions are shown for the PS1xGALEX-SVC sample in Figure~\ref{fig:PS1_corner}, PS1-($i-z$) sample in Figure~\ref{fig:VICK_corner}, and the SDSS spectroscopic sample in Figure~\ref{fig:SDSS_corner}. As a verification of the segmented linear model, we compare the maximized likelihood function with that of a simple linear model. To do so, we consider the Bayesian information criterion (BIC; \citealt{Schwarz:1978}). The BIC takes into account the goodness of fit for each model, while also penalizing the number of free parameters required by each, as a means of mitigating overfitting. This metric is defined as
\begin{equation}
    \textrm{BIC} = -2 \ln{\mathcal{L_{\rm max}}} + dim(\theta)\ln{N},
\end{equation}

\noindent where $\mathcal{L}_{\rm max}$ is the value of the likelihood function corresponding to the optimized parameters, $\theta$, and N is the number of stars in the sample. 

When comparing best-fit models, the preferred model is that exhibits the lowest BIC value. We note that BIC values are only meaningful when compared against models in which the same samples are used. In other words, BIC values for one BHB sample may not be compared to another.

The BIC values for the linear, BIC$_{\rm linear}$, and segmented linear, BIC$_{\rm break}$, models were computed for each of the BHB samples, the results of which are listed in Table~\ref{table:BIC_values}, The resulting posterior distributions are shown for the PS1xGALEX-SVC sample in Figure~\ref{fig:PS1_corner}, PS1-($i-z$) sample in Figure~\ref{fig:VICK_corner}, and the SDSS spectroscopic sample in Figure~\ref{fig:SDSS_corner}. For each model, the difference between the linear and segmented model, $\Delta\textrm{BIC}$, is in excess of $10^2$, where a $\Delta\textrm{BIC}>10$ is considered very strong evidence against the model with the higher BIC. We conclude that, for all three BHB samples considered in this work, the segmented model is a significantly improved representation of the relative age distribution, as compared to a simple linear model.

As an independent verification of the segmented linear model, we performed RANSAC modeling on the inner- and outer-halo regions, inferring a break radius from the intersection of the linear profiles. The derived median age at $R_{G}=0$\,kpc, $A_0$, the break radius, $R_b$, IHR age gradient, $m_{\rm in}$, OHR age gradient, $m_{\rm out}$, and their corresponding uncertainties for the RANSAC method and the maximum likelihood MCMC method are listed in Table~\ref{table:param_values}.

\begin{table}
\label{BIC_table}
	\caption{BIC for Age Models\label{table:BIC_values}}             
	\centering                          
	\begin{tabular}{c | c c c }        
		\hline\hline                  
		BHB Sample - Selection & BIC(linear) & BIC(break) & $\Delta$~BIC \\  
		
		\hline
		PS1xGALEX-SVC        & 51,539 & \textbf{46,865} & 4674\\
        PS1-$(i-z)$          & 63,295 & \textbf{60,245} & 3050\\
        SDSS-Spectroscopic   & 8939  & \textbf{7978}  & 961\\

		\hline        
		
		\hline                                   
	\end{tabular}
\end{table}

The resulting IHR and OHR age gradients, as well as the break radius, vary significantly between the three BHB samples and indeed between the two methods included in this study. Particularly for the PS1xGALEX and PS1-$(i-z)$ sample, this variation is representative of the influence of the photometric selection function on the resulting radial age profile. The purity of the PS1-$(i-z)$ sample was estimated in \citet{Vickers:2012} to be $\sim74$\,\%, with a completeness of $\sim57$\,\%, whereas we demonstrate using the SDSS spectrosopic sample a purity and completeness in excess of 80\% for the PS1xGALEX-SVC sample. Both the PS1xGALEX-SVC sample and the SDSS spectrosopic sample exhibit a larger contrast in their IHR and OHR gradients than the PS1-$(i-z)$ sample. Further, as seen in Figure~\ref{fig:radial_profile}, and in both the MLE and RANSAC determinations in Table~\ref{table:param_values}, the PS1xGALEX values are essentially intermediate to the PS1-$(i-z)$ and SDSS spectroscopic sample estimates. We therefore interpret this as an indication of our improved selection method, and conclude that the purity and completeness BHB samples are essential in order to reveal the contrasting signatures between the IHR and OHR.

For all three BHB samples, the IHR exhibits a significantly steeper age gradient than determined by previous studies that treated the halo as a single profile: $-63.4\pm8.2$\,Myr kpc$^{-1}$, $-44.4 \pm 14.3$\,Myr kpc$^{-1}$, and $-84.3 \pm 14.6$\,Myr kpc$^{-1}$ for the PS1xGALEX-SVC, PS1-$(i-z)$, and SDSS spectroscopic samples, respectively. The gradient in the outer halo of ${\sim-25}$\,Myr kpc$^{-1}$, is consistent with previous studies \citep{Carollo:2016, Das:2016}. The three BHB samples studied in this work represent significantly different selection functions. The unanimity of the contrasting IHR and OHR age gradients, in addition to the superiority of the segmented linear regression model as compared to a simple linear model, supports the conclusion that the break radius seen in the radial age profile is a true signature, as opposed to an artifact of the SVC selection function.

Both the age estimates corrected for metallicity and those inferred from $(g-r)_0$ color for the SDSS spectroscopic sample agree to within $\pm130$\,Myr within $10 < R_{G} < 27$\,kpc, thus the observed gradient in the relative age profile cannot be solely explained by a metallicity gradient in the Milky Way halo. Within $R_{\rm GC} < 10$, Figure\ref{fig:santucci_offset} the photometric age determination is seen to underestimate compared with the age estimates corrected for the spectroscopic metallicity. This suggests that, for the PS1xGALEX-SVC and PS1-$(i-z)$ samples, the IHR gradient is perhaps even steeper than determined in this study.

The difference in the median $(g-r)_0$ color between $5 < R_{G}\textrm{ (kpc)} < 12$ of 0.018\,mag for the PS1xGALEX - SVC sample corresponds to a difference in $(B-V)_0$ of $\sim0.20$\,mag, roughly consistent with the \citet{Preston:1991} result. Both the MLE and RANSAC results are roughly consistent with a break radius occurring at $R_{\rm G} \sim 14$\,kpc, although this value varies somewhat for each sample. The scatter in the IHR and OHR age profiles, $\sigma_{\rm in}$ and $\sigma_{\rm out}$, varies inversely with the sample size of the BHB selection; the largest scatter is seen in the SDSS spectroscopic sample ($N=2,695$), where $\sigma_{\rm out} = 1.48$\,Gyr, while the smallest is seen in the PS1 - $(i-z)$ sample ($N=20,351$), where $\sigma_{\rm in}=1.04$\,Gyr. 

We attribute the larger scatter seen for the SDSS spectroscopic sample to the smaller sample size and comparatively limited sky coverage. The resulting relative age profile is likely to be significantly more susceptible to deviations as a result of substructures in the halo, for instance the Virgo Overdensity \citep{Vivas:2001}. It is reasonable to assume that, as larger samples of BHB stars are obtained from future large-sky surveys, the underlying radial age profile can be further constrained.

\begin{table*}[!htbp]
	\caption{Segmented Linear Profile Parameters for BHB Catalogs\label{table:param_values}}             
	\centering                          
	\begin{tabular}{c | c c c c c c c c c c}        
		\hline\hline                  
		BHB Sample - Selection& $A_0$ & $\sigma_{A_0}$ & $\sigma_{\rm in}$ & $\sigma_{\rm out}$ & $R_b$ & $\sigma_{R_b}$ & $m_{\rm in}$ & $\sigma_{m_{\rm in}}$ & $m_{\rm out}$ & $\sigma_{m_{\rm out}}$\\  
		& (Gyr) & (Gyr) & (Gyr) & (Gyr) & (kpc) & (kpc) & (Myr~kpc$^{-1}$) & (Myr~kpc$^{-1}$) & (Myr~kpc$^{-1}$) & (Myr~kpc$^{-1}$) \\ 
		
		\hline   
		\multicolumn{11}{c}{Maximum likelihood parameters} \\
		\hline
		PS1xGALEX-SVC  & 13.22 & 0.09 & 1.22 & 1.17 & 14.8 & 0.8 & $-$-63.4 & 8.2 & $-$27.2 & 2.5\\
        PS1-$(i-z)$  & 13.12 & 0.13 & 1.04 & 1.07 & 11.0 & 1.3 & $-$44.4 & 14.3 & $-$21.3 & 1.6\\
        SDSS spectroscopic  & 13.19 & 0.17 & 1.43 & 1.52 & 16.2 & 2.4 & $-$84.3 & 14.6 & $-$24.3 & 37.4\\

		\hline
		\multicolumn{11}{c}{Random sample consensus parameters} \\
		\hline
		PS1xGALEX-SVC                & 13.2 & 0.02 & 1.21 & 1.18 & 14.7  & 0.6 & $-$60.9  & 2.1  & $-$27.1 & 1.2 \\ 
		PS1-$(i-z)$     & 13.3 & 0.08 & 1.06 & 1.07 & 9.9  & 1.1 & $-$60.7   & 7.4  & $-$29.3 & 2.6 \\ 
		SDSS spectroscopic  & 13.6 & 0.05 & 1.45 & 1.48 & 13.2  & 0.6 & $-$119.5 & 6.6  & $-$22.5 & 4.9 \\ 
		\hline        
		
		\hline                                   
	\end{tabular}
\end{table*}

\section{Interpretation}
We compare our results to the Aquarius halo Aq-C-5 \citep{Tissera:2013}, which best reproduced the outer-halo age profile in \citet{Carollo:2018}, for which our RANSAC method produces an IHR gradient of $-30.0 \pm 1.8$\,Myr kpc$^{-1}$. This value is somewhat smaller than our observational result of $-63.4 \pm 8.2$\,Myr kpc$^{-1}$, though a value of $-48.6 \pm 2.5$\,Myr kpc$^{-1}$ is obtained when considering only the accreted component of Aq-C-5. Considering that the median metallicity for the SDSS spectroscopic sample of [Fe/H]$=-1.75$ most closely resembles the median metallicity of the accreted component of Aq-C-5 \citep[see][for details.]{Tissera:2013}, as opposed to the in situ component, it is possible that BHB stars selected in the manner outlined by this work are a natural probe of the halo's accreted population. This is consistent with the result from \citet{Carollo:2018}, where the slope of the age profile is largely determined by the accreted stellar populations acquired in early stages of halo assembly. Complementary to these studies, \citet{Fernandez:2019} showed that the more massive accreted satellites in Aq-C-5 (and Aq-D-5) have extended star-formation activity, consistent with being gas-rich galaxies.

The contrasting age gradients between the inner and outer regions of the stellar halo may therefore reflect the contrasting roles of dissipational -- i.e. gas-rich -- and dissipationless mergers in the assembly histories of the halo components. In the dissipative merger scenario, star formation in gas-rich satellites can continue throughout the accretion event. Alternatively, where dissipationless mergers occur, we expect a halo dominated by stars donated from smaller satellite galaxies with truncated star-formation histories \citep{Chiba:2000, Carollo:2007, Carollo:2010}. The break radius seen in the radial age profile could therefore represent the transition from an IHR dominated by dissipative mergers, to an OHR characterized by dissipationless accretion of lower mass sub-galactic fragments.

An open question is whether the ancient inner-halo structure proposed by \citet{Helmi:2018} constitutes both the high-eccentricity component -- the \textit{Gaia} Sausage -- previously identified by \citet{Belokurov:2018}, and the retrograde component of the halo, or if these structures represent distinct progenitor systems, as evidenced by \citet{Myeong:2019}. For an age gradient arising purely through accretion, according to the hierarchical process described in \citet{Amorisco:2017}, our results favor two or more distinct progenitors of the inner halo. The proposed \textit{Gaia}-Enceladus is thought to have undergone $\sim2$\,Gyr of star formation \citep{Helmi:2018}. Considering the estimated infall time of $\sim 10$\,Gyr ago, with the oldest members stars being $13$\,Gyr, \textit{Gaia}-Enceladus likely wasn't forming stars by the infall time. Unless an extended star formation occurred in \textit{Gaia}-Enceladus, or \textit{Gaia}-Enceladus possessed a significant radial age gradient prior to accretion, the age gradient in BHB stars seems to rule out it being the single progenitor of the inner halo.

As suggested by \citet{Deason:2018}, the break radius seen in the stellar density profile at $R_G \sim 20$\,kpc \citep{Sesar:2011} can be explained as the result of coincident apocentric radii of stars in the highly eccentric component of the halo. However, the break radius seen in the relative age profile of BHB stars occurs at a significantly smaller Galactocentric radius, $R_b \sim 14$\,kpc. Considering the uncertainty in the stellar density break radius of $\pm1$\,kpc \citep{Xue:2015}, and the uncertainty in our determination of the break radius in the relative age profile of $\pm2$\,kpc, we do not expect the break radius of the stellar density profile to be associated with the break radius in the relative age profile, and thus this discrepancy can be explained if at least two populations with characteristically distinct ages inhabit the region within the apocentric radius of the \textit{Gaia} Sausage.

Alternatively, for an inner halo assembled from only a handful of relatively massive progenitor systems \citep{Helmi:2018, Myeong:2018, Kruijssen:2019, Lancaster:2019}, it is feasible that these systems retained sufficiently high gas-to-star ratios to enable persistent star formation throughout the merger event. For an age gradient that is not instantiated by the accretion, and is thus driven by in situ star formation, we speculate that the steepness of the radial age profile might reflect the dynamics of the star-formation history, mass-assembly rate, or gas contributions from the progenitor systems, provided that the number of progenitors is small. These gas-rich mergers would then be subject to ram pressure stripping during their interactions with the host halo \citep{Simpson:2018}. While this pressure has the capability to remove cold gas from the satellite and quench star formation, the hot virialized gas in the host galaxy can act to shield the dwarf from both ram-pressure stripping and UV-background heating.
If we assume the infall of a few ($N\sim3$) progenitors, as suggested from recent observational and simulation studies, then the spread in the BHB age distribution in the IHR of $\sim1.0$\,Gyr suggests either (1) a persistent star formation throughout the merger process, effectively constraining the thermal to ram pressure ratio \citep{Hausammann:2019} of the mergers that contributed to the formation of the inner halo, or (2) a significant contrast between the ages of the stellar populations contributed to the halo by the Gaia Sausage, Sequoia, and possibly additional dwarf systems yet to be identified.


\section{Conclusions}

Using selections of BHB stars from Pan-STARRS DR1 and GALEX photometry, we demonstrate the first evidence of a break radius in the relative age profile of the Milky Way, occurring at $R_{\textrm{GC}}\sim 14$\,kpc. Within the break radius, we measure a significantly steeper age gradient than previous studies, $-63.4 + \pm 8.2$\,Myr kpc$^{-1}$. A novel methodology was developed for the selection of BHB stars from photometry, using 3D support vector classification. We demonstrate, using the catalog of spectroscopically confirmed BHBs from \citet{Santucci:2015b}, that we have achieved an unprecedented selection purity, $\sim 80$\,\%. Age distributions inferred from photometry are offset slightly from those determined from BHB population synthesis, which corrects for influence of metallicity, but otherwise preserve the gradient signature, verifying the use of photometric colors as a reasonable approximation of relative age for BHB samples.

Our results confirm that $(g-r)_0$ gradient in BHB stars corresponds to a negative age gradient consistent with the ``inside-out" formation model, wherein the oldest halo stars populate the inner-halo region. The contrasting age gradients in the inner- and outer-halo region suggest that the inner and outer haloes have fundamentally different formation histories. We postulate that the steeper age gradient seen in the inner-halo region is evidence of the dissipational formation of the inner halo, consisting of a few massive progenitor systems. The existence of a break radius and unique age gradients in the inner- and outer-halo regions provide additional constraints for simulations of galactic formation, particularly for the inferred mass-assembly and merger-tree histories of the Milky Way.

\acknowledgments

The authors thank the referee for their comments, which substantially improved our analysis.
D.D.W., T.C.B., V.M.P., and P.D. acknowledge partial support for this work from grant PHY 14-30152; Physics Frontier Center/JINA Center for the Evolution of the Elements (JINA-CEE), awarded by the US National Science Foundation. T.C.B. acknowledges partial support from the Leverhulme Trust (UK), which hosted his visiting professorship at the University of Hull during the completion of this study.

R.M.S. gratefully acknowledges Conselho Nacional de Desenvolvimento Científico e Tecnológico (CNPq, process No 436696/2018-5).

The Pan-STARRS1 Surveys (PS1) and the PS1 public science archive have been made possible through contributions by the Institute for Astronomy, the University of Hawaii, the Pan-STARRS Project Office, the Max Planck Society and its participating institutes, the Max Planck Institute for Astronomy, Heidelberg and the Max Planck Institute for Extraterrestrial Physics, Garching, The Johns Hopkins University, Durham University, the University of Edinburgh, the Queen's University Belfast, the Harvard-Smithsonian Center for Astrophysics, the Las Cumbres Observatory Global Telescope Network Incorporated, the National Central University of Taiwan, the Space Telescope Science Institute, the National Aeronautics and Space Administration under grant No. NNX08AR22G issued through the Planetary Science Division of the NASA Science Mission Directorate, the National Science Foundation grant No. AST-1238877, the University of Maryland, Eotvos Lorand University (ELTE), the Los Alamos National Laboratory, and the Gordon and Betty Moore Foundation.

Funding for the Sloan Digital Sky Survey IV has been provided by the Alfred P. Sloan Foundation, the U.S. Department of Energy Office of Science, and the Participating Institutions. SDSS-IV acknowledges
support and resources from the Center for High-Performance Computing at
the University of Utah. The SDSS website is www.sdss.org.

SDSS-IV is managed by the Astrophysical Research Consortium for the 
Participating Institutions of the SDSS Collaboration, including the 
Brazilian Participation Group, the Carnegie Institution for Science, 
Carnegie Mellon University, the Chilean Participation Group, the French Participation Group, Harvard-Smithsonian Center for Astrophysics, 
Instituto de Astrof\'isica de Canarias, The Johns Hopkins University, Kavli Institute for the Physics and Mathematics of the Universe (IPMU) / 
University of Tokyo, the Korean Participation Group, Lawrence Berkeley National Laboratory, 
Leibniz Institut f\"ur Astrophysik Potsdam (AIP),  
Max Planck Institut f\"ur Astronomie (MPIA Heidelberg), 
Max Planck Institut f\"ur Astrophysik (MPA Garching), 
Max Planck Institut f\"ur Extraterrestrische Physik (MPE), 
National Astronomical Observatories of China, New Mexico State University, 
New York University, University of Notre Dame, 
Observat\'ario Nacional / MCTI, The Ohio State University, 
Pennsylvania State University, Shanghai Astronomical Observatory, 
United Kingdom Participation Group,
Universidad Nacional Aut\'onoma de M\'exico, University of Arizona, 
University of Colorado Boulder, University of Oxford, University of Portsmouth, 
University of Utah, University of Virginia, University of Washington, University of Wisconsin, 
Vanderbilt University, and Yale University.

\newpage
\appendix

\begin{figure*}
	\label{fig:XZ_final}
	\begin{center}
		\includegraphics[trim = 0.20cm 0.50cm 0.75cm 0.50cm, clip, width=\columnwidth]{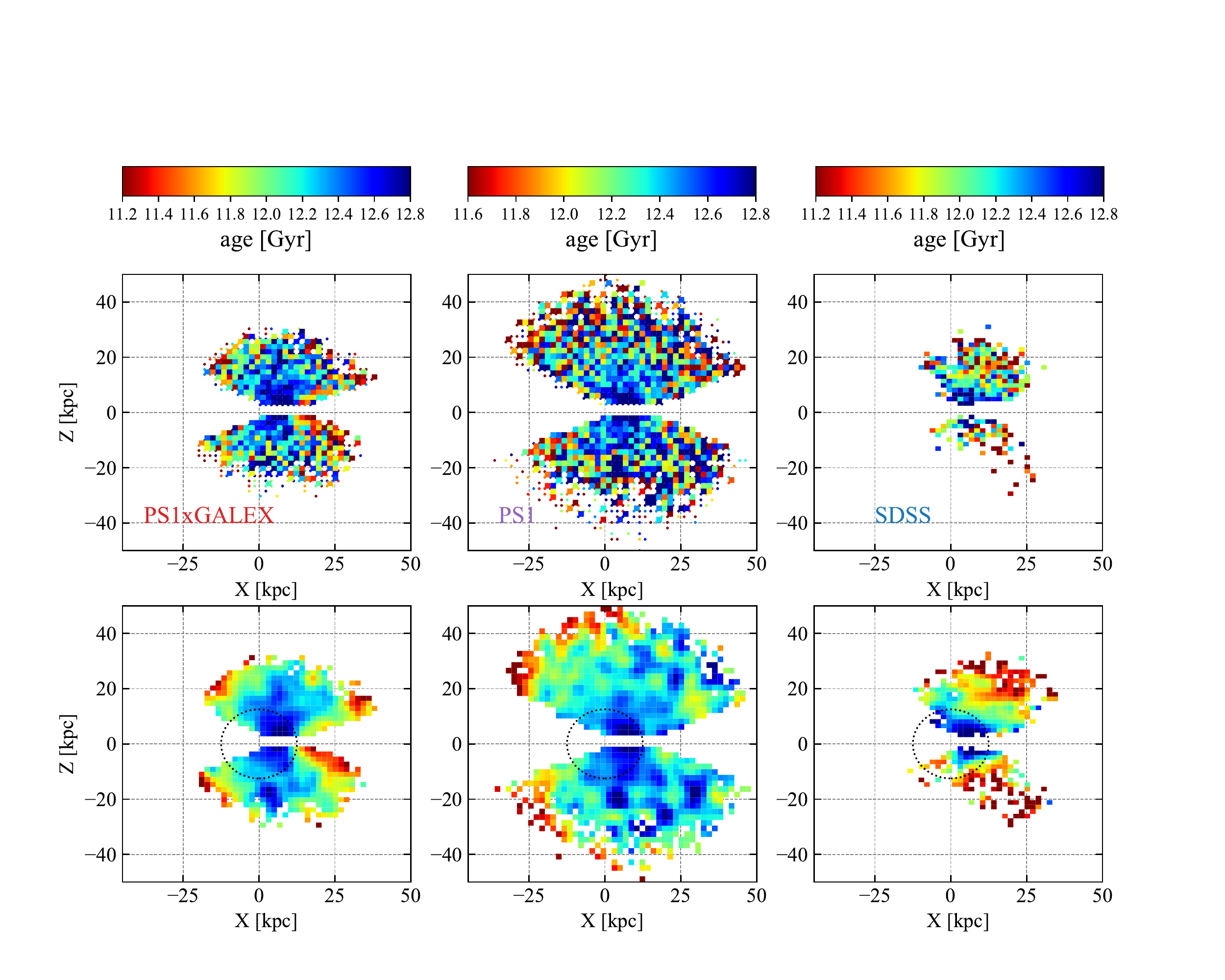}
		
		\caption{Projections of Galactocentric X versus Z for the PS1xGALEX (\textit{left}), PS1-$(i-z)$ (\textit{center}), and SDSS spectroscopic (\textit{right}) samples. The dotted circle represents the approximate transition from the IHR to the OHR, $R_{G} =  12.5$\,kpc. }
	\end{center}
\end{figure*}

\begin{figure*}
	\label{fig:YZ_final}
	\begin{center}
		\includegraphics[trim = 0.20cm 0.50cm 0.75cm 0.50cm, clip, width=\columnwidth]{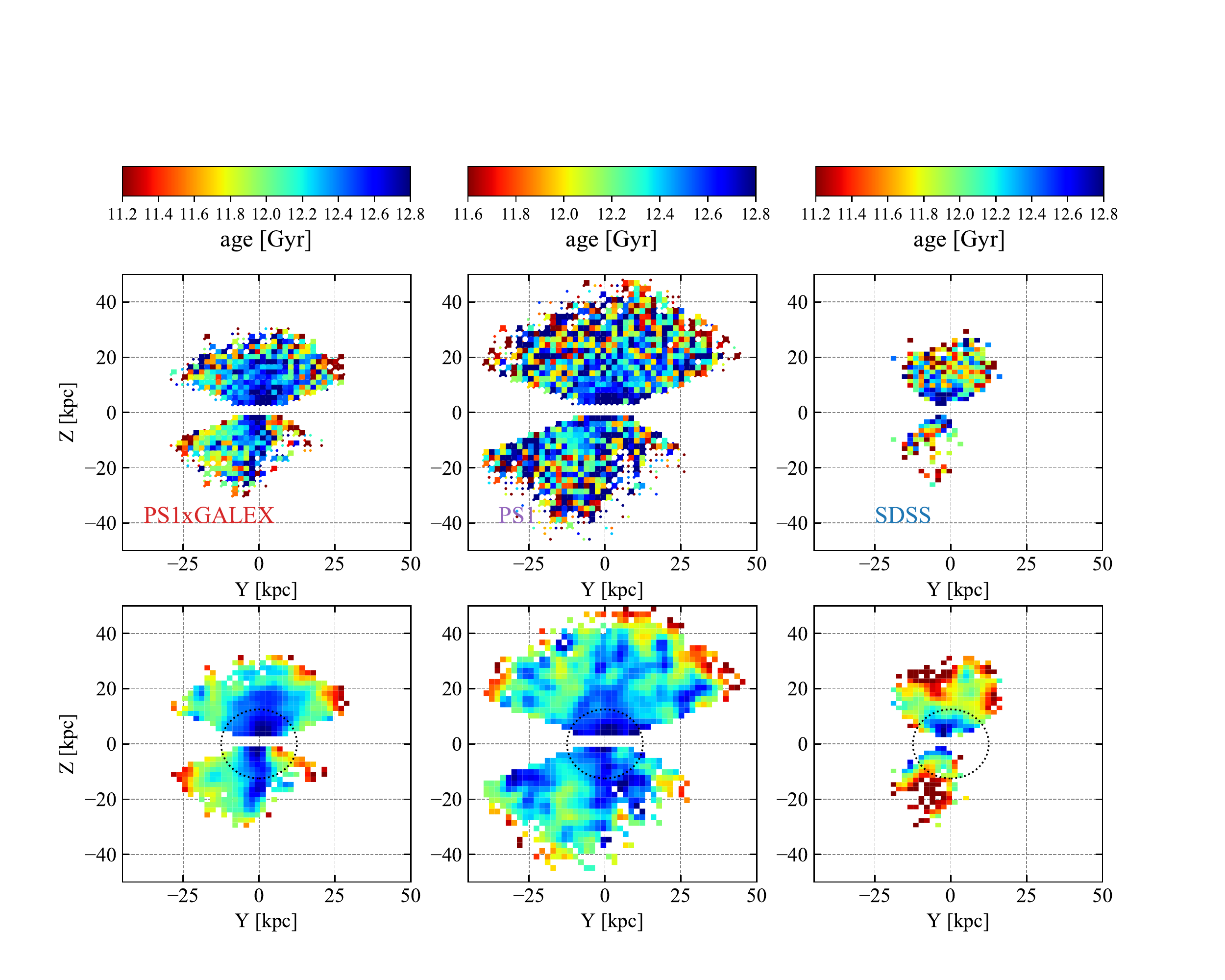}
		
		\caption{Projections of Galactocentric Y vs. Z for the PS1xGALEX (\textit{left}), PS1 - $(i-z)$ (\textit{center}), and SDSS spectroscopic (\textit{right}) samples. The dotted circle represents the approximate transition from the IHR to the OHR, $R_{G} =  12.5$\,kpc. }
	\end{center}
\end{figure*}

\begin{figure*}
	\label{fig:XY_final}
	\begin{center}
		\includegraphics[trim = 0.20cm 0.50cm 0.75cm 0.50cm, clip, width=\columnwidth]{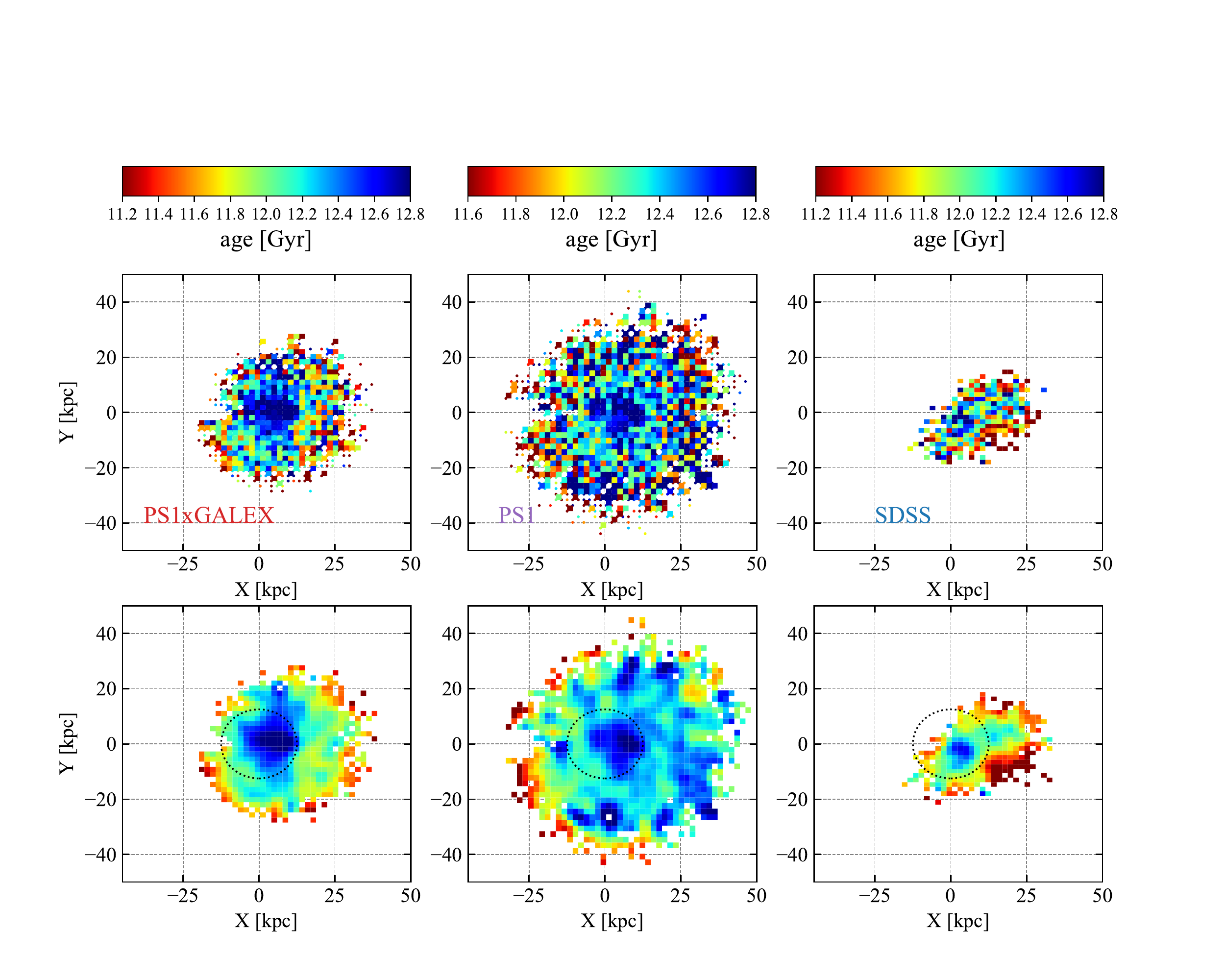}
		
		\caption{Projections of Galactocentric X vs. Y for the PS1xGALEX (\textit{left}), PS1-$(i-z)$ (\textit{center}), and SDSS spectroscopic (\textit{right}) samples. The dotted circle represents the approximate transition from the IHR to the OHR, $R_{G} =  12.5$\,kpc. }
	\end{center}
\end{figure*}

In this section, we present the relative age projections in Galactocentric X versus Z (Figure~\ref{fig:XZ_final}), Y versus Z (Figure~\ref{fig:YZ_final}), and X versus Y (Figure~\ref{fig:XY_final}) coordinates, for each of the BHB samples described in the text. We provide the X versus Z age distributions of the SDSS spectroscopic sample from \citep{Santucci:2015a} in Figure 10, for both photometric estimates and corrected estimates from \citet{Denissenkov:2017}. We present the resulting posterior probability distributions from the MLE for each the PS1xGALEX - SVC sample (Figure~\ref{fig:PS1_corner}), the PS1-($i-z$) sample (Figure~\ref{fig:VICK_corner}), and the SDSS spectroscopic sample (Figure~\ref{fig:SDSS_corner}).
\clearpage

\begin{figure*}
\includegraphics[trim = 1.0cm 1.0cm 2.0cm 3.0cm, width=\textwidth]{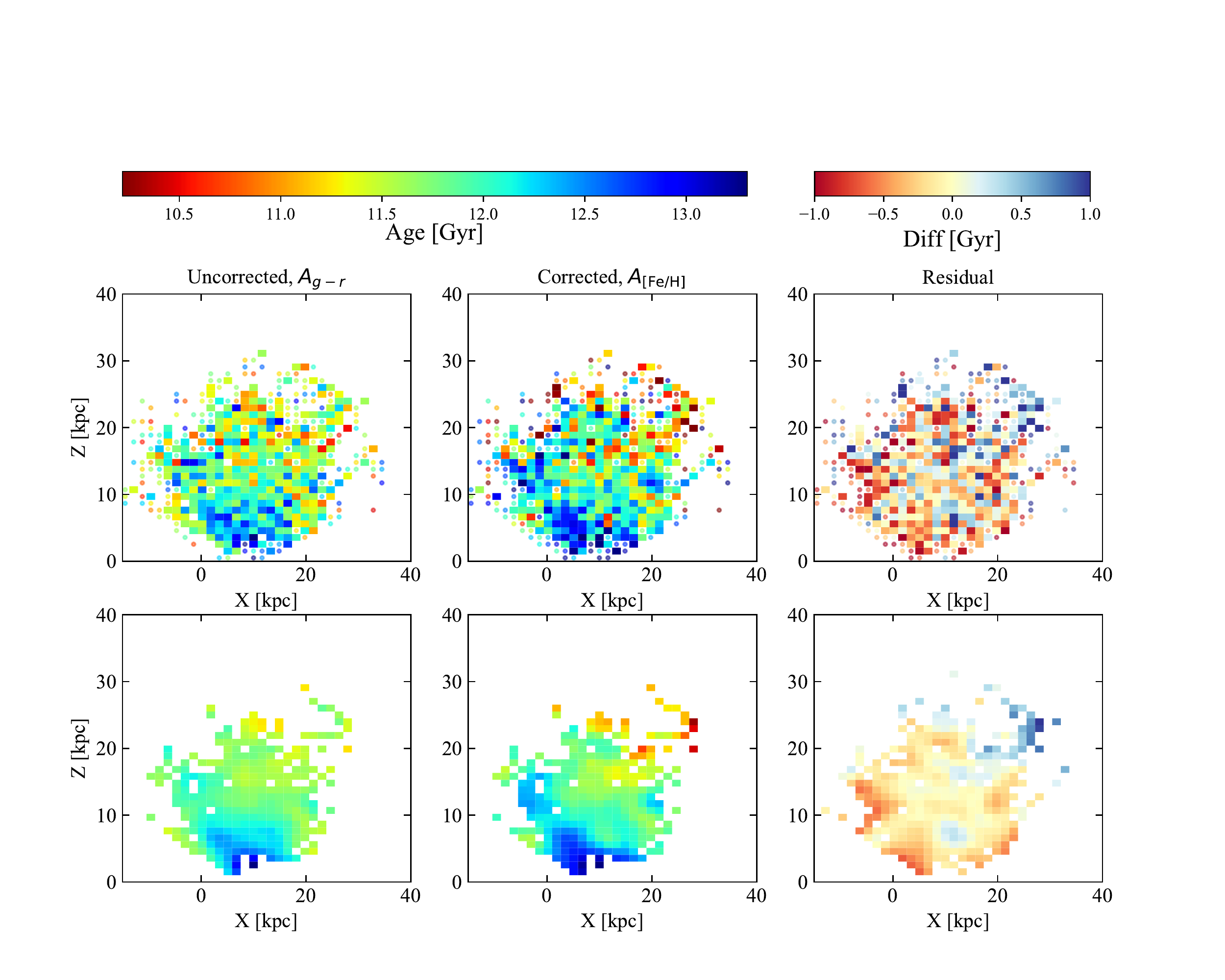}
\caption{Age distributions of the SDSS spectroscopic sample from \citet{Santucci:2015b}, for both photometric estimates and corrected estimates from \citet{Denissenkov:2017}.}
	\label{fig:santucci_map}
\end{figure*}

\begin{figure*}
	\includegraphics[trim = 3.0cm 0.00cm 3.0cm 0.0cm, width=\textwidth]{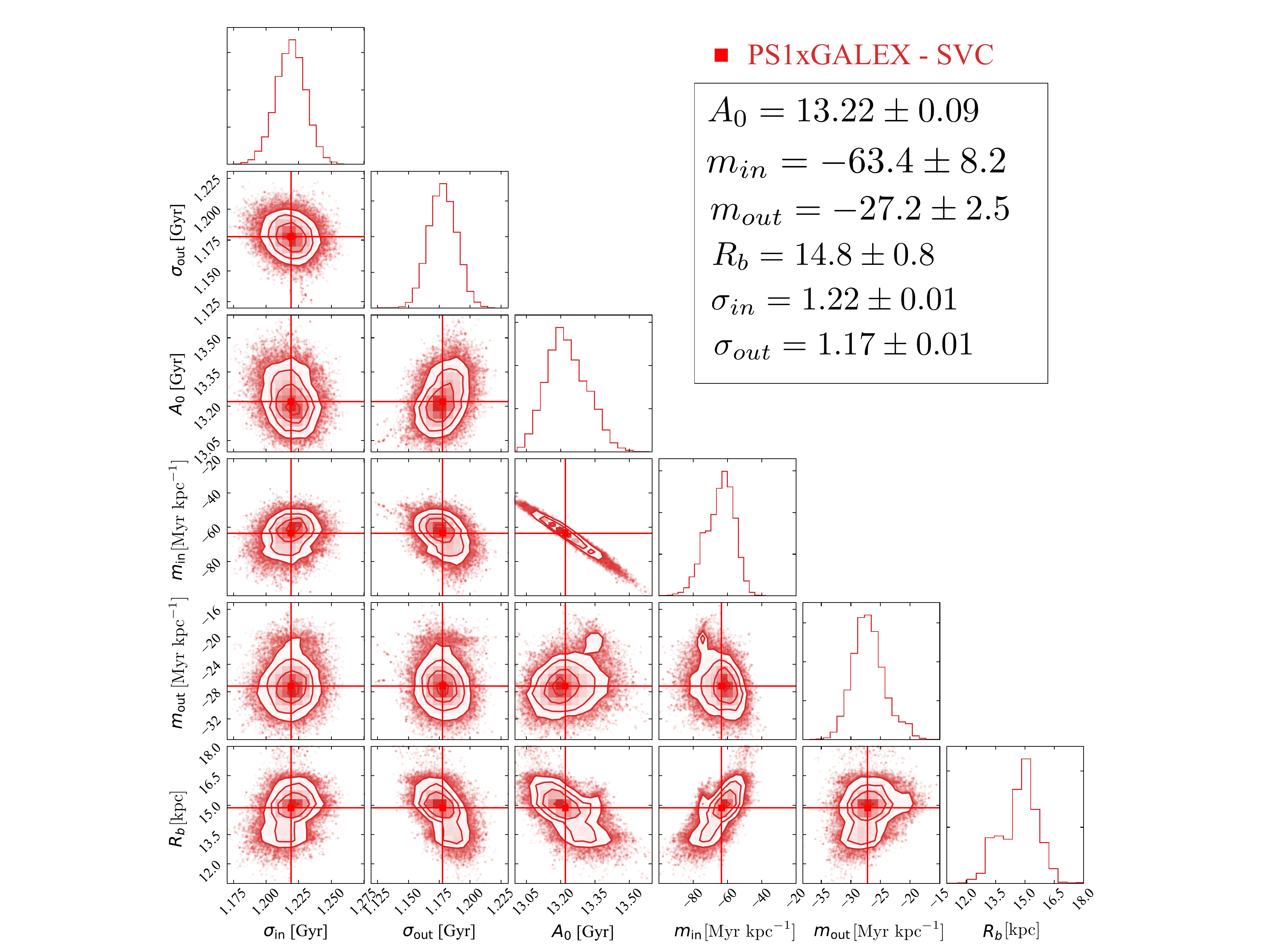}
	\caption{Posterior probability distributions of parameters ($\sigma_{\rm in}$, $\sigma_{\rm out}$, $A_0$, $m_{\rm in}$, $m_{\rm out}$, $R_b$) of the segmented linear regression model for the PS1xGALEX-SVC sample. The red lines and squares correspond to the most likely values.}
	\label{fig:PS1_corner}
\end{figure*}

\begin{figure*}
	\includegraphics[trim = 3.0cm 0.00cm 3.0cm 0.0cm, width=\textwidth]{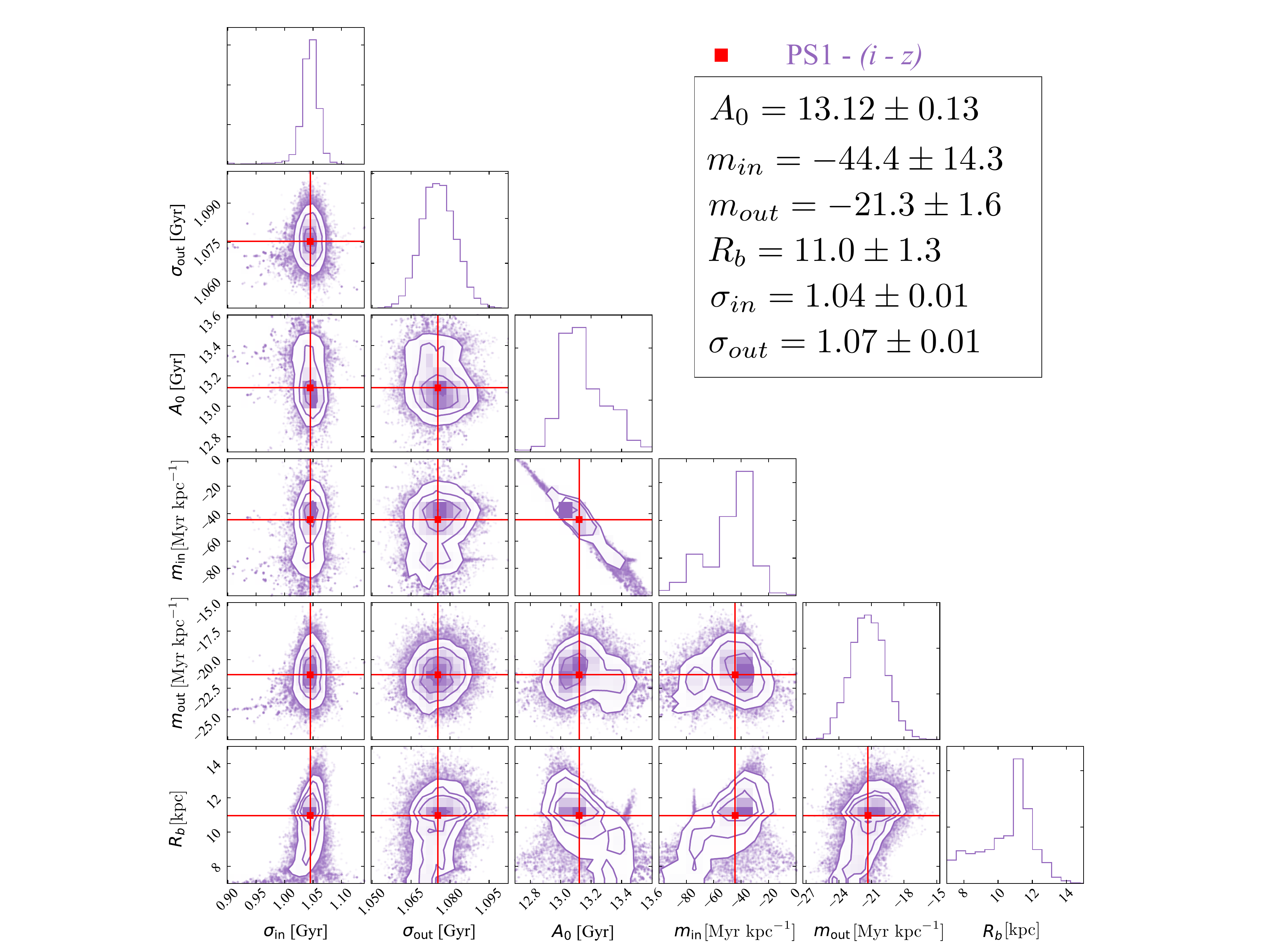}
	\caption{Posterior probability distributions of parameters ($\sigma_{\rm in}$, $\sigma_{\rm out}$, $A_0$, $m_{\rm in}$, $m_{\rm out}$, $R_b$) of the segmented linear regression model for the PS1-$(i-z)$ sample. The red lines and squares correspond to the most likely values.}
	\label{fig:VICK_corner}
\end{figure*}

\begin{figure*}
	\includegraphics[trim = 3.0cm 0.00cm 3.0cm 0.0cm, width=\textwidth]{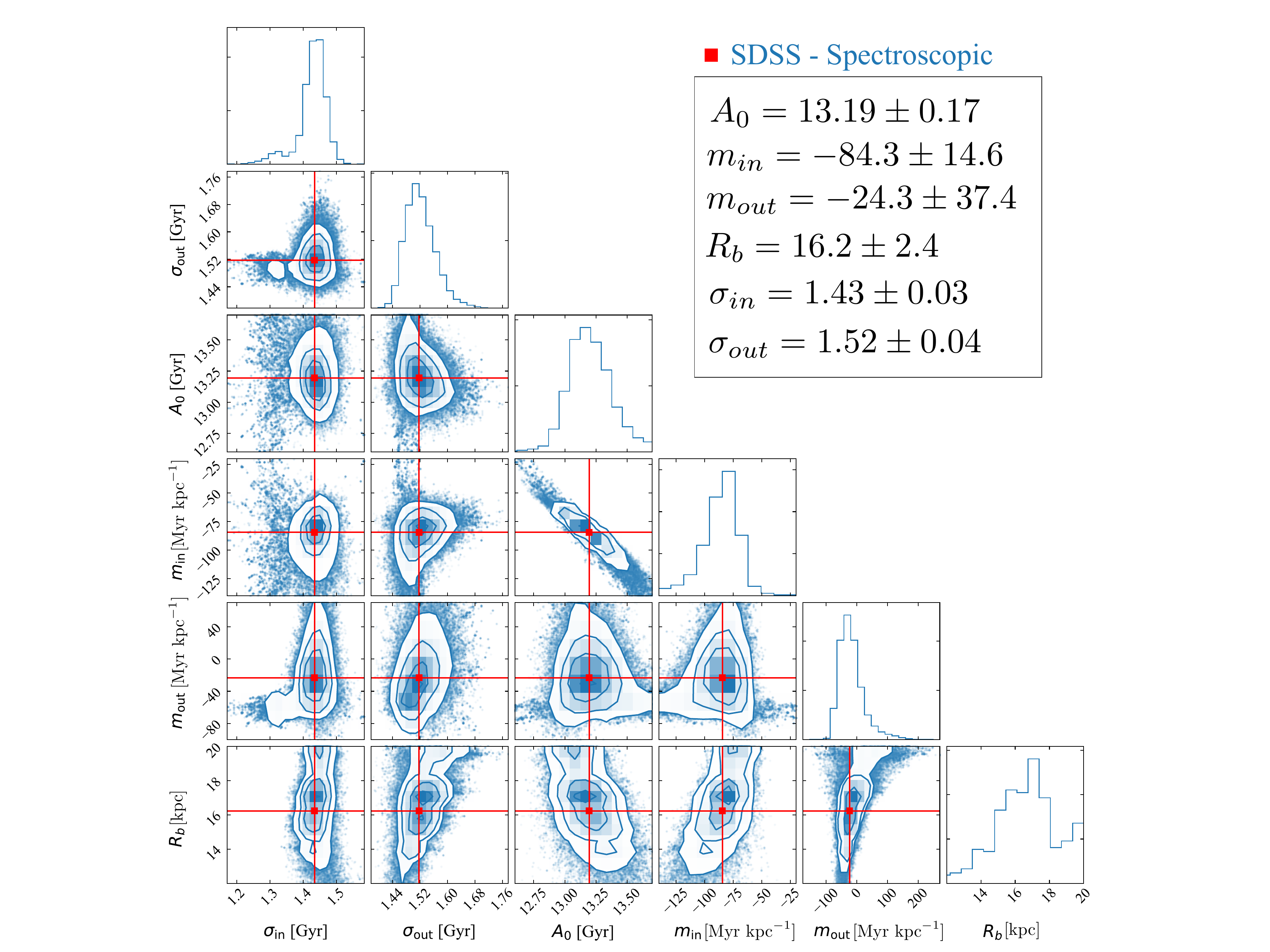}
	\caption{Posterior probability distributions of parameters ($\sigma_{\rm in}$, $\sigma_{\rm out}$, $A_0$, $m_{\rm in}$, $m_{\rm out}$, $R_b$) of the segmented linear regression model for the SDSS spectroscopic sample. The red lines and squares correspond to the most likely values.}
	\label{fig:SDSS_corner}
\end{figure*}

\clearpage

\bibliographystyle{yahapj}
\nocite{*}
\bibliography{BHB.bib}

\end{document}